\documentclass[ashowpacs,preprintnumbers,superscriptaddress,aps,prd,nofootinbib,floatfix,11pt]{revtex4-2}
\usepackage[T1]{fontenc}
\usepackage[utf8]{inputenc}
\usepackage{amsmath,amssymb}
\usepackage{booktabs}
\usepackage{multirow}
\usepackage{graphicx}
\usepackage{float}
\usepackage{placeins} 
\usepackage{lmodern}
\usepackage{microtype}
\usepackage{xurl}
\usepackage{url}
\usepackage{hyperref}

\usepackage{dcolumn}
\usepackage{bm}
\usepackage{physics,amsfonts}
\usepackage{latexsym}
\usepackage{color}
\usepackage{caption}
\usepackage{ulem}
\usepackage{mathrsfs}
\usepackage{yhmath}
\usepackage{cancel}
\usepackage{tensor}
\usepackage[table]{xcolor}
\usepackage{colortbl}
\usepackage{array}
\usepackage[usestackEOL]{stackengine}

\usepackage{titlesec}
\renewcommand{\thesection}{\arabic{section}}
\renewcommand{\thesubsection}{\arabic{subsection}}
\hypersetup{%
setpagesize=false,
 bookmarksnumbered=true,%
 bookmarksopen=true,%
 colorlinks=true,%
 linkcolor=blue,
 citecolor=magenta,
 urlcolor=blue
}

\def\nn    {\nonumber}
\DeclareUnicodeCharacter{039B}{\Lambda}

\newcolumntype{P}[1]{>{\centering\arraybackslash}p{#1}}
\newcolumntype{Q}[1]{>{\raggedright\arraybackslash}p{#1}}

\definecolor{colHdr}{RGB}{60,100,170}
\definecolor{colSub}{RGB}{210,222,242}
\definecolor{colA}{RGB}{244,247,253}
\definecolor{colB}{RGB}{255,255,255}
\definecolor{colGrp}{RGB}{225,232,245}
\definecolor{colCvg}{RGB}{240,243,250}

\def\apre{a_{\rm pre}}

\def\tpre{T_{\rm pre}}
\def\t0{T_0}

\def\he{H_{\rm e}}

\def\omegax{\Omega_{\chi}h^2}
\def\mx{m_{\chi}}

\titleformat{\section}
  {\normalfont\large\bfseries}{\thesection}{1em}{}
\titlespacing*{\section}
  {0pt}{1.ex plus .1ex minus .1ex}{0.ex plus .1ex minus .1ex}

\makeatletter
\renewcommand{\p@subsection}{\thesection.}
\renewcommand{\p@subsubsection}{\thesection.\thesubsection.}
\makeatother

\titleformat{\subsubsection}
  {\normalfont\itshape}{\thesection.\thesubsection.\thesubsubsection}{1em}{}
\titlespacing*{\subsubsection}
  {0pt}{1.ex plus .1ex minus .1ex}{0.ex plus .1ex minus .1ex}

\renewcommand{\thesubsection}{.\arabic{subsection}}

\begin{document}
\allowdisplaybreaks
\flushbottom

\title{\boldmath Gravitational Dark Matter Production from Spectator Field in $R^2$-Higgs Inflation}

\author{Arnab Samanta}
\affiliation{Department of Physical Sciences, Indian Institute of Science Education and Research Berhampur, Berhampur 760003, Odisha, India}
\author{Rajesh Mondal}
\affiliation{Department of Physical Sciences, Indian Institute of Science Education and Research Berhampur, Berhampur 760003, Odisha, India}
\author{Tanmoy Modak}
\affiliation{Department of Physical Sciences, Indian Institute of Science Education and Research Berhampur, Berhampur 760003, Odisha, India}

\begin{abstract}
We investigate the cosmological gravitational dark matter production of spectator field $\chi$ in the multifield $R^2$-Higgs inflation. The spectator field $\chi$ couples to gravity only through a non-minimal coupling $\xi_\chi\chi^2R$ and serves as a dark matter (DM) candidate. Working consistently in the Einstein frame, we derive the equation of motion of the background fields and fluctuations at linear order using the doubly covariant formalism.  We show the dark matter production and the subsequent comoving relic abundance heavily depend on this effective mass of the field $\chi$. Specifically, we find that there exists cancellation regions in the effective mass-squared of $\chi$, which ensures the correct relic abundance is achieved. Further, we show that the preheating from Higgs and Goldstone may play critical role in accounting the observed relic density.
\end{abstract}

\maketitle

\vspace{2ex}

\hrule
\vspace{2ex}
 \tableofcontents
\vspace{2ex}
\numberwithin{equation}{section}
\hrule
\setlength{\parskip}{1\baselineskip}
\setlength{\parindent}{0pt}
\vspace{1ex}

\section{Introduction}
\label{sec:intro}
Dark matter constitutes a fundamental component of the Universe, yet its direct detection continues to elude us. While various of astrophysical and cosmological observations such as galaxy rotation curve, cosmic microwave background anisotropy, and gravitational lensing have provided tantalizing evidence for the existence of dark matter, the production process and its true nature are still unknown. Among the various proposed production mechanisms, the nonperturbative gravitational production due to time-dependent background is particularly intriguing within the context of early Universe physics primarily because it requires no coupling between dark matter and the Standard Model (SM) sector~\cite{Ford:1986sy}. Instead, the dark matter is generated purely through the expansion of spacetime in the evolving Universe.

In the nonperturbative gravitational DM production, the production is induced primarily due to the time dependence of the background metric associated with cosmic expansion. A spectator field coupled nonminimally with gravity in such time-dependent background may account for the observed DM relic density ~\cite{Chung:1998ua,Chung:1998zb,Chung:2001cb,Chung:2018ayg,Ema:2018ucl,Kolb:2023ydq,Ling:2021zlj,Garcia:2022vwm,Garcia:2023qab,Jenks:2024fiu,Racco:2024aac,Verner:2024agh,Chung:2011ck,Li:2019ves,Pineda:2024olh,Kolb:2026xfp,Watanabe:2015eia,Kolb:2022eyn,Laverda:2026slq,Chakraborty:2025zgx}. In case of both single and multifield inflation, if such a spectator field coupled with the time dependent inflationary background, the DM particle can be produced nonperturbatively after the end of inflation~\cite{Chung:1998ua,Chung:2001cb,Chung:2018ayg,Ema:2018ucl,Kolb:2023ydq,Ling:2021zlj,Garcia:2022vwm,Garcia:2023qab,Jenks:2024fiu,Racco:2024aac,Verner:2024agh,Chung:2011ck,Li:2019ves}.

We investigate the gravitational production of DM from a spectator scalar field $\chi$, nonminimally coupled to gravity through $\xi_\chi \chi^2 R$, in $R^2$-Higgs inflation. Existing studies often examine spectator particle production in the Jordan frame while describing inflationary dynamics in the Einstein frame. Although this approximation may yield reasonable estimates of the relic abundance, it can obscure important contributions to particle production. In multifield models such as $R^2$-Higgs inflation, the Weyl transformation induces additional contributions to the spectator's time-dependent effective mass, arising from the transformed scalar potential and the curved field-space geometry. Consequently, the effective mass depends not only on the mass parameter $m_\chi$ and nonminimal coupling $\xi_\chi$, but also on the inflationary parameters $\xi_R$, $\xi_H$, and $\lambda$, where $\xi_R$ is the Ricci self coupling associated with $R^2$ term, $\xi_H$ is the Higgs nonminimal coupling and $\lambda$ is the Higgs quartic coupling. Employing a doubly covariant formalism, we demonstrate how these additional contributions influence gravitational particle production and the resulting DM relic abundance.

A second common approximation in estimating the DM relic abundance is to treat the reheating temperature as a free parameter. However, in $R^2$-Higgs inflation, efficient Higgs and Goldstone preheating can lead to thermalization even for $\xi_H \gtrsim 1$~\cite{Cado:2024von,Modak:2025bjv,Gonuguntla:2026rkw}. Consequently, when thermalization proceeds through preheating, the resulting temperature cannot be arbitrarily adjusted but is instead determined by the underlying inflationary dynamics. We show that the parameter space that accounts for the DM relic density induces the correct preheating temperature $\tpre$ needed to account for the observed DM relic density. Further, at the end of preheating, the linear-order approximation typically used to estimate the DM abundance from a spectator field breaks down. Therefore, the comoving DM number density should reach a constant value, i.e., freeze, well before preheating is complete. Preheating therefore serves two key purposes: (i) for a specified particle content, it determines the preheating temperature and hence $\Omega_\chi h^2$; (ii) it establishes the time frame within which the comoving number density must reach an asymptotic plateau for linear theory to remain valid. Thus, estimating $\tpre$ rather than treating it as a free parameter is essential for the validity of linear theory. We show that these requirements somewhat tightly constrains the parameter space for gravitational DM production from a spectator field.

The paper is organized as follows. Sec.~\ref{sec:seeaction} presents the Jordan- and Einstein-frame actions and background equations. Sec.~\ref{sec:inflation} discusses background and perturbation dynamics while in Sec.~\ref{sec:spectator} we examine the gravitational production of $\chi$, its number density and relic abundance, and the viable parameter space. In Sec.~\ref{sec:prehiggs} we study the Higgs and Goldstone preheating and its impact on the relic abundance. We summarize our results with some discussion in Sec.~\ref{sec:discu}.

\section{Action and dynamics}
\label{sec:seeaction}
The action of $R^2$-Higgs inflation with a spectator scalar $\chi$ nonminimally coupled to gravity in the Jordan frame is given by
\begin{align}
  S_J  =  \int d^4 x \sqrt{-g_J} \bigg[ \frac{M_{\rm P}^{2}}{2} f(R_J, \Phi)
  -g_J^{\mu\nu}(\nabla_\mu\Phi)^\dagger \nabla_\nu\Phi-g_J^{\mu\nu}\partial_\mu\chi\ \partial_\nu\chi  - V_\Phi(\Phi, \Phi^\dagger)-V_\chi(\chi)\nonumber\\ -  \dfrac{1}{4} g_J^{\mu\rho} g_J^{\nu\sigma} B_{\mu\nu}B_{\rho\sigma}
  - \dfrac{1}{4} g_J^{\mu\rho} g_J^{\nu\sigma} W^i_{\mu\nu}W^i_{\rho\sigma}
 \bigg],\label{eq:actionJ1}
\end{align}
where,
\begin{align}
f(R_J, \Phi,\chi) &= R_J  + \frac{\xi_R}{2 M_{\rm P}^2} R_J^2   + \frac{2\,\xi_H}{M_{\rm P}^2} |\Phi|^2 R_J-\frac{\xi_\chi}{M_{\rm P}^2} \chi^2 R_J, \\\ V_\Phi(\Phi, \Phi^\dagger) &= \lambda|\Phi|^4\,,V_\chi(\chi)=\frac{1}{2}\,m^2_\chi\,\chi^2\label{def:fRterm},
\end{align}
where, $M_{\rm P}=\sqrt{1/\left(8\pi G\right)}\approx 2.4\times 10^{18}~\text{GeV}$ is the reduced Planck mass with $G$ being Newton's gravitational constant. The chosen metric convention is $(-1,+1,+1,+1)$ and $\sqrt{-g_J}$ denotes the  determinant of the metric and $R_J$ is the space-time Ricci scalar in Jordan frame. The $\Phi$ is the Higgs field with hypercharge $+1$. The bare Higgs mass term is dropped since it has negligible impact in the inflationary dynamics. The spectator field $\chi$, of mass $m_\chi$, does not acquire a background value and is treated throughout as a quantum fluctuation; its only tree-level coupling to the gravitational sector is via $\xi_\chi$.
 $B_{\mu\nu}$ and $W^i_{\mu\nu}$ are the field-stress tensors of the SM gauge groups respectively.
The covariant derivative for the $SU(2)_L \times U(1)_Y$ groups are defined as
\begin{equation}
\nabla_\mu = D_\mu + i g' \frac{1}{2} Q_{Y_f} B_\mu + i g \, \bm{T} \cdot \bm{W}_\mu,
\end{equation}
where $g'$ and $g$ respective gauge couplings, $\bm{T}={\bm{\tau}}/{2}$ are the weak-isospin and $Q_{Y_f}$ is the $U(1)_Y$ hypercharge with,  $\bm{\tau}$ are Pauli matrices. Note that the field-stress tensors of gauge fields are
\begin{align}
B_{\mu\nu} = D_\mu B_\nu - D_\nu B_\mu, \hspace{1.8cm} W^i_{\mu\nu} = D_\mu W^i_\nu - D_\nu W^i_\mu - g \sum_{j,k=1}^{3} \epsilon_{ijk} W^j_\mu W^k_\nu,
\end{align}

In general, analyzing the dynamics of $R^2$-Higgs inflation is computationally more convenient when formulated in the Einstein frame.
To transform the  generic $f(R_J, \Phi,\chi)$ theory action in Eq.~\eqref{eq:actionJ1}
to the scalar-tensor theory we first introduce an auxiliary field $\Psi$ and perform a Legendre transformation.  The action is rewritten as
\begin{equation} \begin{aligned}
  S_J  =  \int d^4 x \sqrt{-g_J}  \bigg[&\frac{M_{P}^{2}}{2} \left(f(\Psi,  \Phi, \chi)
  + \frac{\partial f(\Psi,  \Phi,\chi)}{\partial \Psi} (R_J-\Psi)\right)
   -g_J^{\mu\nu}(\partial_\mu\Phi)^\dagger \partial_\nu\Phi -g_J^{\mu\nu}\partial_\mu\chi\ \partial_\nu\chi\\& - V_\Phi(\Phi, \Phi^\dagger)-V_\chi(\chi)-  \dfrac{1}{4} g_J^{\mu\rho} g_J^{\nu\sigma} B_{\mu\nu}B_{\rho\sigma}
   - \dfrac{1}{4} g_J^{\mu\rho} g_J^{\nu\sigma} W^i_{\mu\nu}W^i_{\rho\sigma}\bigg]\label{eq:actionJ2}.
\end{aligned} \end{equation}

The Legendre transformation remains well-defined provided that $f(R_J,\Phi,\chi)$ exhibits strict convexity~\cite{Odintsov:2017hbk}. Consequently, it is very convenient to introduce the associated physical degree of freedom by introducing
\begin{align}
\Theta = \frac{\partial f(\Psi,  \Phi,\chi)}{\partial \Psi},\label{eq:theta}
\end{align}
and rewrite the action in Eq.~\eqref{eq:actionJ2} as
\begin{equation} \begin{aligned}
  S_J  =  \int d^4 x \sqrt{-g_J} \bigg[& \frac{M_{P}^{2}}{2} \Theta R_J - U(\Theta,   \Phi)
  -g_J^{\mu\nu}(\partial_\mu\Phi)^\dagger \partial_\nu\Phi- g_J^{\mu\nu}\partial_\mu\chi\ \partial_\nu\chi - V_\Phi(\Phi, \Phi^\dagger)-V_\chi(\chi) \bigg].\label{eq:actionJ4}
\end{aligned} \end{equation}
The solution of Eq.~\eqref{eq:theta} can be written as,
\begin{align}
\Psi = \frac{M_{\rm P}^2}{\xi_R}\left(\Theta-1- \frac{2\,\xi_H}{M_{\rm P}^2} |\Phi|^2+\frac{\xi_\chi}{M_{\rm P}^2} \chi^2\right)
\end{align}
The potential $U(\Theta,\Phi,\chi)$ with degrees of freedom $\Theta$, $\Phi$ and $\chi$ takes the form
\begin{align}
U(\Theta,   \Phi,\chi) &= \frac{M_{\rm P}^2}{2}\left[\Psi\, \Theta -  f(\Psi,  \Phi,\chi)\right] = \frac{M_{\rm P}^4}{4\,\xi_R }\left(\Theta-1- \frac{2\,\xi_H}{M_{\rm P}^2} |\Phi|^2+\frac{\xi_\chi}{M_{\rm P}^2} \chi^2\right)^2
\end{align}

With the field redefinition and decomposition of Higgs field (with $Q_Y=+1$)
\begin{align}
\phi = M_{\rm P} \sqrt{\frac{3}{2}} \ln\Theta,~~~
\Phi =
\frac{1}{\sqrt{2}} \begin{pmatrix}
  \phi_3+ i \phi_4 \\
  h + i \phi_2 \\
\end{pmatrix},
\end{align}
the action at quadratic order in Einstein frame is
\begin{equation} \begin{aligned}
S_E  &= \int d^4 x \sqrt{-g_E}\bigg[\frac{M_{\rm P}^2}{2} R_E - \frac{1}{2} G_{IJ} g_E^{\mu \nu} D_\mu \phi^I D_\nu \phi^J -
V_E(\phi^I) -  \dfrac{1}{4} g_E^{\mu\rho} g_E^{\nu\sigma} F_{A\mu\nu}F_{A\rho\sigma}-
\dfrac{1}{4} g_E^{\mu\rho} g_E^{\nu\sigma} F_{Z\mu\nu}F_{Z\rho\sigma} \\
&  - \dfrac{1}{2} g_E^{\mu\rho} g_E^{\nu\sigma} F^+_{W\mu\nu}F^-_{W\rho\sigma} - e^{-\sqrt{\frac{2}{3}}\frac{\phi}{M_{\rm P}}} g_E^{\mu\nu}\bigg(\frac{g_Z^2}{8}h^2 Z_\mu Z_\nu+
\frac{g_Z}{2}\left[ (D_\mu h) \ \phi_2- (D_\mu\phi_2) h\right] Z_\nu +\frac{e^2}{4 s_W^2} h^2  W^+_\mu W^-_\nu+\\
&\frac{i e}{2 \sqrt{2} s_W} D_\mu h \left[W^-_\nu  (\phi_3+i \phi_4)  - W^+_\nu   (\phi_3-i \phi_4) \right]-
\frac{i e}{2 \sqrt{2} s_W}  \left[ W^-_\nu D_\mu(\phi_3+i\phi_4)  -  W^+_\nu D_\mu(\phi_3-i\phi_4) \right] h\bigg]\label{eq:actionfinal},
\end{aligned} \end{equation}
where  $g_E^{\mu \nu}$ is the space-time metric in Einstein frame and $g_Z= {e}/({s_W c_W})$, with $s_W$ denotes the sine of the Weinberg angle.
Here the field space is spanned by fields $\phi^I \in \{\phi, h,\chi, \phi_2, \phi_3, \phi_4 \}$. The  $G_{IJ}$s are the $6\times 6$ field-space metric   with non-vanishing elements $G_{\phi\phi} = 1$, $G_{ h h} = e^{-\sqrt{\frac{2}{3}}\frac{\phi}{M_{\rm P}}}$,$G_{\chi\chi} = e^{-\sqrt{\frac{2}{3}}\frac{\phi}{M_{\rm P}}}$,$G_{\phi_i \phi_i} = e^{-\sqrt{\frac{2}{3}}\frac{\phi}{M_{\rm P}}}$ with $i=2,3,4$.
The potential $V_E$ is given as
\begin{align}
V_E(\phi^I) = &e^{-2\sqrt{\frac{2}{3}}\frac{\,\phi}{M_{\rm P}}}\Biggl[\frac{\lambda}{4} \left(h^2+\sum^4_{i=2} \phi_i^2\right)^2+\frac{1}{2}m_\chi^2\,\chi^2\nonumber\\&+\frac{M_{\rm P}^4}{4\,\xi_R }\left(e^{\sqrt{\frac{2}{3}}\frac{\phi}{M_{\rm P}}}-1- \frac{\xi_H}{M_{\rm P}^2} \left(h^2+\sum^4_{i=2} \phi_i^2\right)^2+\frac{\xi_\chi}{M_{\rm P}^2} \chi^2\right)^2\Bigg].
\label{eq:potential}
\end{align}
We have treated the massive scalar field $\chi$ as well as goldstones $\phi_i$'s as perturbations, i.e., they do not acquire any background values.
The potential $V_E(\phi^I)$ reads ,
\begin{equation}
    V_E(\phi^I)\simeq
V_0(\phi^I)+\frac{1}{2}e^{-2\sqrt{\frac{2}{3}}\frac{\,\phi}{M_{\rm P}}}\left(m_\chi^2+\frac{M_P^2\,\xi_\chi}{\xi_R}e^{\sqrt{\frac{2}{3}}\frac{\phi}{M_{\rm P}}}F(\phi^I)\right)\chi^2,
\end{equation}
where,
\begin{equation}
\begin{aligned}
    &V_0(\phi^I)=\frac{\lambda}{4}e^{-2\,\alpha\,\phi}\,\left(h^2+\sum_{i=2}^{4}\phi_i^2\right)^2+\frac{M_P^4}{4\,\xi_R}F^2(\phi^I)\,,\\
&F(\phi^I)=1-e^{-\alpha\,\phi}-e^{-\alpha\,\phi}\frac{\xi_H}{M_{\rm P}^2}\left(h^2+\sum_{i=2}^{4}\phi_i^2\right)\,,\\
&\alpha=\sqrt{\frac{2}{3}}\frac{1}{M_{\rm P}}\,.
\end{aligned}
\end{equation}
Varying the quadratic action we get the equations of motion (EoMs) for the scalars fields $\phi^I$ as
\begin{equation} \begin{aligned}
&\Box \phi^K + \Gamma^{K}_{\ IJ} \ g_E^{\alpha \nu} D_\alpha  \phi^I D_\nu \phi^J - G^{KM} V_{E,M} \\&+ G^{KM}
e^{-\sqrt{\frac{2}{3}}\frac{\phi}{M_{\rm P}}} \left(\sqrt{\frac{2}{3}}\frac{1}{M_{\rm P}}\right) g_E^{\alpha\nu} (\partial_\alpha \phi)
\bigg(\frac{g_Z}{2} \delta^3_M h Z_\nu +\frac{i e}{2 \sqrt{2} s_W}  \left[ W^-_\nu  (\delta^4_M+i\delta^5_M)  -  W^+_\nu  (\delta^4_M-i\delta^5_M) \right] h\bigg)\\
&-G^{KM}e^{-\sqrt{\frac{2}{3}}\frac{\phi}{M_{\rm P}}} g_E^{\alpha\nu}
\bigg(\frac{g_Z}{2}   \ D_\alpha\left(\delta^3_M h Z_\nu\right)
+\frac{i e}{2 \sqrt{2} s_W}  D_\alpha\left[\left( W^-_\nu  (\delta^4_M+i\delta^5_M)  -  W^+_\nu  (\delta^4_M-i\delta^5_M) \right] h\right)\bigg)  \\
&-G^{KM}\bigg[e^{-\sqrt{\frac{2}{3}}\frac{\phi}{M_{\rm P}}} g_E^{\mu\nu}\delta^3_M\bigg(\frac{g_Z}{2}(D_\mu h) Z_\nu\bigg)+e^{-\sqrt{\frac{2}{3}}\frac{\phi}{M_{\rm P}}} g_E^{\mu\nu}\delta^4_M\bigg(\frac{i e}{2 \sqrt{2} s_W} D_\mu h \left(\ W^-_\nu - W^+_\nu \right)\bigg)\\
&+e^{-\sqrt{\frac{2}{3}}\frac{\phi}{M_{\rm P}}} g_E^{\mu\nu}\delta^5_M\bigg(\frac{i e}{2 \sqrt{2} s_W} D_\mu h \left(\ i W^-_\nu + i W^+_\nu \right)\bigg)\bigg]
= 0\label{eom:scalar},
\end{aligned} \end{equation}
where $\Gamma^{K}_{\ IJ}$ are the field-space Christoffel symbols. Note here that, we have only considered action at quadratic
order for all fields in Eq.~\eqref{eq:actionfinal} since we restrict ourselves to linear approximation.

\section{Dynamics of Inflation}
\label{sec:inflation}
To investigate the inflationary dynamics, as well as perturbation dynamics, the fields $\phi^I(x^\mu)$ are decomposed into homogeneous background part (${\varphi}^I $) and perturbation ($\delta\phi^I$) as
\begin{align}
\phi^I(x^\mu) = \varphi^I(t) + \delta\phi^I(x^\mu)\label{fieldexpan},
\end{align}
where $t$ is the cosmic time. Here, only $\phi$ and $h$ acquire background field values i.e., $\varphi^I(t) = \{\varphi(t),h_0(t)\}$.
The presence of multiple scalar fields coupled with non-canonical kinetic terms motivates us to follow the doubly covariant formalism~\cite{Gong:2011uw,Kaiser:2012ak,Sfakianakis:2018lzf} at the linear order in perturbation. The linear order perturbed Friedmann-Robertson-Walker (FRW) metric is given as~\cite{Kodama:1984ziu,Mukhanov:1990me,Malik:2008im}
\begin{align}
ds^2 &= -(1+2\mathcal{A}) dt^2 +2 a(t) (\partial_i \mathcal{B}) dx^i dt +
a(t)^2 \left[(1-2\psi) \delta_{ij}+ 2 \partial_i \partial_j \mathcal{E}\right] dx^i dx^j,\label{eq:frwmetric}
\end{align}
where $\mathcal{A}, \mathcal{B}, \psi$ and $\mathcal{E}$ are the metric scalar perturbations and $a(t)$
is scale factor. Throughout this work we consider the longitudinal gauge where the scalar perturbations $\mathcal{E}$ and $ \mathcal{B}$ vanish.
At the linear order of perturbation, one can simply find the EoMs of the background fields from Eq.~\eqref{eom:scalar}
\begin{align}
&\mathcal{D}_t \dot{\varphi}^I + 3 H\dot{\varphi}^I + G^{I J} V_{E,J}= 0\label{eq:bkg_inf},
\end{align}
where $_{,J}$ denotes derivative with respect to fields, $\mathcal{D}_t$ and  $\mathcal{D}_J$ are field space covariant derivatives defined as~\cite{Gong:2011uw,Sfakianakis:2018lzf,Kaiser:2012ak}
\begin{align}
\mathcal{D}_t A^I   = \dot{A}^I  + \Gamma^I_{\; JK} \dot{\varphi}^J A^K,\,\mathcal{D}_J A^I  = \partial_J A^I + \Gamma^I_{\; JK} A^K,
\end{align}
Here, in Eq.~\eqref{eq:bkg_inf}, the $G^{I J}$, $\Gamma^I_{\; JK}$ and $V_{E,J}$ are evaluated at the background order.

The Hubble parameter is defined as
\begin{align}
H^2 &= \left(\frac{\dot{a}}{a}\right)^2 = \frac{1}{3 M_{\rm P}^2} \bigg(\frac{1}{2} G_{IJ} \dot{\varphi}^I \dot{\varphi}^J + V_0(\varphi^I)\bigg),\label{hubble1}\\
\dot{H} &= -\frac{1}{2 M_{\rm P}^2} \bigg(G_{IJ} \dot{\varphi}^I \dot{\varphi}^J\bigg).\label{hubble2}
\end{align}
The slow-roll parameter $\epsilon$ and the number of $e$-folding $\mathcal{N}$ are expressed as
\begin{align}
\epsilon = -\frac{\dot{H}}{H^2},
~~~\mathcal{N}(t) \equiv \ln \frac{a(t)}{a(t_{e})},\label{epsiefol}
\end{align}
where $t_{e}$ is the corresponding cosmic time when the inflation ends and  $a(t_{e})$ is the scale factor at the end of inflation.
We denote the end of inflation when $\epsilon (t=t_{e}) = 1$.
In what follows, we shall use the $t$ and $\mathcal{N}$ interchangeably.
The energy density associated with background fields is
\begin{align}
&\rho_{\mathrm{inf}} = \frac{1}{2} G_{IJ} \dot{\varphi}^I \dot{\varphi}^J + V_E(\varphi^I),\label{eq:infenergy}
\end{align}
where $G_{IJ}$ and $V_E$ are evaluated at the background field order.

In order to understand the production of the spectator field $\chi$, we shall now focus on the EoMs of the field fluctuations $\delta\phi^I(x^\mu)$.
In general $\delta\phi^I(x^\mu)$ are gauge-dependent quantities but one can conveniently construct the covariant field fluctuations
$\mathcal{Q}^I(x^\mu)$ which relates the $\phi^I(x^\mu)$
to corresponding $\varphi^I(t)$ via a unique geodesic in the field-space where
$\delta\phi^I$ is expressed as~\cite{Gong:2011uw,Elliston:2012ab}
\begin{align}
\delta\phi^I &= \mathcal{Q}^I -\frac{1}{2} \Gamma^I_{\ JK} \mathcal{Q}^K \mathcal{Q}^J+\frac{1}{3!} \big(\Gamma^I_{\ MN} \Gamma^N_{\ JK}-\Gamma^I_{\ JK,M}\big)  \mathcal{Q}^K \mathcal{Q}^J  \mathcal{Q}^M+\dots~.
\end{align}
At the linear order $\mathcal{Q}^I = \delta\phi^I$ and the gauge-independent Mukhanov-Sasaki variables
are~\cite{Mukhanov:1990me,Sasaki:1986hm,Mukhanov:1988jd}
\begin{align}
Q^I = \mathcal{Q}^I + \frac{\dot{\varphi}^I}{H}\psi = \delta\phi^I+ \frac{\dot{\varphi^I}}{H}\psi.\label{mukh-sasaki}
\end{align}
Note that is $Q^I$s are doubly covariant with respect both the space-time and field-space transformations.
In the field-space manifold $\dot{\varphi}^I$ and $Q^I$ transform like vectors.
We may now insert Eq.~\eqref{mukh-sasaki} and Eq.~\eqref{eq:frwmetric} into Eq.~\eqref{eom:scalar} to find
EoMs for $Q^I$s at linear order as~\cite{Sfakianakis:2018lzf,Cado:2024von,Modak:2025grj},
\begin{equation}
 \mathcal{D}_t^2 Q^{I} + 3 H \mathcal{D}_t Q^{I} -\frac{\partial^2}{a^2} Q^{I} + \mathcal{M}^{I}_{\ \ J}\, Q^{J}+ \mathcal{A}_{(\phi_I)}= 0,\label{eq:perturb}
\end{equation}
where $\mathcal{A}_{(\phi)}=\mathcal{A}_{(h)}=\mathcal{A}_{(\chi)}=0\,,$  $\mathcal{A}_{(\phi_2)}= g_Z [(1/\sqrt{6}M_{\rm P})\,\dot{\phi_0} h_0 Z_0-\dot{h}_0 Z_0+ (h_0/2)(D_\nu Z^\nu)]$, and $g_Z\to i e/(2\sqrt{2}s_W)$ and, $Z_0\to (W^-_0-W^+_0)$ and $(iW^-_0+iW^+_0)$ for $\mathcal{A}_{(\phi_3)}$ and $\mathcal{A}_{(\phi_4)}$, respectively. As only $\phi$ and $h$ acquire background values, the EoMs of $Q^{\chi}$, $Q^{\phi_2}$, $Q^{\phi_3}$ and $Q^{\phi_4}$ decouple from $Q^{\phi}$ and $Q^{h}$.
The the mass-squared matrix $\mathcal{M}^{I}_{\ L}$ can be written as,
\begin{align}
\mathcal{M}^{I}_{\ L} = G^{IJ} (\mathcal{D}_L\mathcal{D}_J V_0)- \mathcal{R}^I_{\ JKL} \dot{\varphi}^J \dot{\varphi}^K
- \frac{1}{M_{\rm P}^2\, a^3} \mathcal{D}_t \left(\frac{a^3}{H}\dot{\varphi}^I \dot{\varphi}_L\right),\label{eq:massterm}
\end{align}
where the $\mathcal{R}^I_{\ JKL}$ is the field-space Riemann tensor
which is evaluated, along with $\mathcal{M}^{I}_{\ L}$, at the background order. The field-space metric $G_{IJ}$ and $\mathcal M_{IJ}$ are diagonal at
background order, so the equations of motion for the first-order perturbations do not mix the different $Q^I$.


\section{Cosmological Gravitational Production of Spectator Field $\chi$}
\label{sec:spectator}
In this section, we discuss the cosmological gravitational production of a massive spectator
scalar field $\chi$ in the multifield inflationary background. It is clear from Eq.~\eqref{eq:actionJ1}, that the field $\chi$ interacts only
gravitationally and may serve as a DM candidate. In the Einstein frame, the presence of a time-dependent effective mass for the spectator field consequently leads to gravitational particle production.

As discussed above, the $\chi$ field do not acquire background field values and, as a result, it decouples from $Q^\phi$ and $Q^h$. From Eq.~\eqref{eq:perturb},
we find the EoM for the fluctuation $Q^\chi$ as
\begin{equation}
    \mathcal{D}_t^2 Q^\chi+ 3 H \mathcal{D}_t Q^\chi -\frac{\nabla^2}{a^2} Q^\chi + \mathcal{M}^\chi_{~~\chi}\, Q^\chi = 0
\end{equation}
In order to quantize the field $\chi$, we move to conformal time $d\tau = dt/a(t)$ and perform field transformation $X^\chi(x^\mu)=a(t)\,Q^\chi(x^\mu)$. The EoM of the field fluctuation $X^\chi$ in conformal time found as
\begin{equation}
    \mathcal{D}_\tau^2 X^\chi -\bigg[\partial^2  - a^2 \left(\mathcal{M}^\chi_{\ \ \chi} -\frac{1}{6}R_E\, G^\chi_{\ \ \chi} \right)\bigg] X^\chi = 0,\label{eqpertur:chi}
\end{equation}
The quantization of the rescaled fluctuations in momentum space can be written as,
\begin{equation}
    \hat{\widetilde{X}}^\chi
=
z_k(\tau)e^{\chi}(\tau)\hat{a}(\mathbf{k})
+
z_k^*(\tau)e^{\chi}(\tau)
\hat{a}^{\dagger}(-\mathbf{k}).
\end{equation}
where $\hat{a}$ and $\hat{a}^{\dagger}$ are respectively annihilation and creation operators. The $z_k$ is the mode function, and
$e^{\chi}$ is the corresponding vielbein,
satisfying the relations $e^{\chi}\,e^{\chi}=G^{\chi\chi}$ with $\mathcal{D}_\tau e^\chi(\tau)=0$. Using the above quantization relation,
the equation of motion for the mode function $z_k$ becomes,
\begin{eqnarray}{\label{dmmode}}
z''_k+\omega_{(\chi)}^2(\tau,k)\,z_k=0\,,
\end{eqnarray}
where the time-dependent frequency $\omega^2_{(\chi)}(\tau,k) =\left(k^2 + a^2 m_{\mathrm{eff},(\chi)}^2(\tau)  \right)$ with the
different components of the effective mass $m_{\mathrm{eff},(\chi)}^2$  is given as
\begin{subequations}
\begin{align}
m_{1,(\chi)}^2 &=G^{\chi J} (\mathcal{D}_{\chi}\mathcal{D}_J V_E)=e^{-\alpha\varphi}\,m_\chi^2+\mathcal{F}_m\label{def:m1}\\
        &\quad \quad \quad \quad ~\mbox{with}~ \mathcal{F}_m=\left[\frac{\xi_\chi\,\rm M_{\rm P}^2}{\xi_R}\,F(\varphi^I)+\alpha^2V_0(\varphi^I)-\frac{\alpha^2\,\rm M_P^4}{4\,\xi_R}F(\varphi^I)\right]
       \nn\\
m_{2,(\chi)}^2 &=- \mathcal{R}^{\chi}_{\ \ JK\chi} \dot{\varphi}^J \dot{\varphi}^K=-\frac{1}{6}\left(e^{-\alpha\phi}\dot{h}_0^2+\dot{\varphi}^2\right)\,,\\
m_{3,(\chi)}^2 &=- \frac{1}{M_{\rm P}^2 a^3} \mathcal{D}_t \left(\frac{a^3}{H}\dot{\varphi}^{\chi} \dot{\varphi}_{\chi}\right)=0\\
m_{4,(\chi)}^2 &=-\frac{R_E}{6}=-(2H^2+\dot H)\label{def:m4}
\end{align}
\end{subequations}
such that $m_{\mathrm{eff},(\chi)}^2=\sum_i m_{i,(\chi)}^2$. The time dependence of the effective mass $m_{\mathrm{eff},(\chi)}^2$ leads to a time-varying frequency $\omega_{(\chi)}(\tau,k)$ in Eq.~\eqref{dmmode}, whose nonadiabatic evolution gives rise to particle production. We remark that some contributions to $m_{\mathrm{eff},(\chi)}^2$ have been overlooked in the literature due to treating DM production and inflationary dynamics in different frames. As we shall see, a delicate cancellation among these terms is crucial for obtaining the correct relic abundance.

The growth of the mode functions $z_k$ leads to an increase in the occupation number $n_k$ of the produced particles. The comoving number density of particles $n_k$ with comoving momentum mode $k$ can be evaluated as the energy
of that mode,
$\frac{1}{2}\left|{z}^\prime_k\right|^2
+
\frac{1}{2}\omega_{(\chi)}^2(\tau,k)\,\left|z_k\right|^2,$
divided by the energy $\omega_{(\chi)}(\tau,k)$ of each particle \cite{Kofman:1997yn}:

\begin{equation}\label{eqnk}
n_k
=
\frac{\omega_{(\chi)}(\tau,k)}{2}
\left(
\frac{|{z}^\prime_k|^2}{\omega_{(\chi)}^2(\tau,k)}
+|z_k|^2
\right)=\frac{a^2(t)\,|\dot{z}_k|^2}{2\,\omega_{(\chi)}(t,k)}+\frac{\omega_{(\chi)}(t,k)\,|z_k|^2}{2}
\end{equation}
The second equality in Eq.~\eqref{eqnk}, we have written in cosmic time $t$, since we solve the mode equation in cosmic time, as given in Eq.~\eqref{dmmode}.
The total comoving energy density can be obtained by integrating the above equation over the momentum \cite{Kofman:1997yn},
\begin{equation}
n_\chi(t)\,a^3(t)=\int\frac{d^3k}{(2\,\pi)^3}n_k(t)\,.
\end{equation}
Here, $n_\chi(t)$ denotes the physical number density of $\chi$ particles. Finally, the vacuum-subtracted comoving number densities associated with the produced particles are given by,
\begin{equation}
Y_\chi(t)=n_\chi(t)\,a^3(t)-n^{\rm BD}_\chi(t)\,a^3(t)=\int\frac{d^3k}{(2\,\pi)^3}\left(n_k(t)-\frac{1}{2}\right)\,,~~~~n^{\rm BD}_\chi(t)\,a^3(t)=\int\frac{k^2}{4\,\pi^2}\,dk\label{eq:ychi}
\end{equation}

\begin{table}[h]
  \centering
  \begin{tabular}{|c|c|c|c|c|c| c|}
    \hline
    BP & $\lambda$ & $\xi_R$ & $\xi_H$  &$\varphi(t_{\rm in})$ $[M_{\rm P}]$& $h_0(t_{\rm in})$ $[M_{\rm P}]$ & $H_e$ $[M_{\rm P}]$\\

    \hline
    $a$ & $10^{-2}$ & $2.13 \times 10^9$ & $100$ &5.37&$7.0\times10^{-6}$ & $3.02\times10^{-6}$ \\
    \hline
     $b$ & $10^{-2}$ & $2.2 \times 10^9$ & $10$ &5.50&$5.0\times10^{-3}$ & $2.97\times10^{-6}$\\
     \hline
    $c$ & $10^{-2}$ & $2.55 \times 10^9$ & $1$ & $5.60$&$8.94\times10^{-4}$ &  $2.76\times10^{-6}$\\
    \hline
  \end{tabular}
  \caption{BPs chosen for our analysis. The scales are given in unit of the Planck mass $M_{\rm P}$.}
  \label{table:bp}
\end{table}

\begin{figure}[h]
 \centering
\captionsetup{justification=raggedright,singlelinecheck=false}
\includegraphics[width=.32 \textwidth]{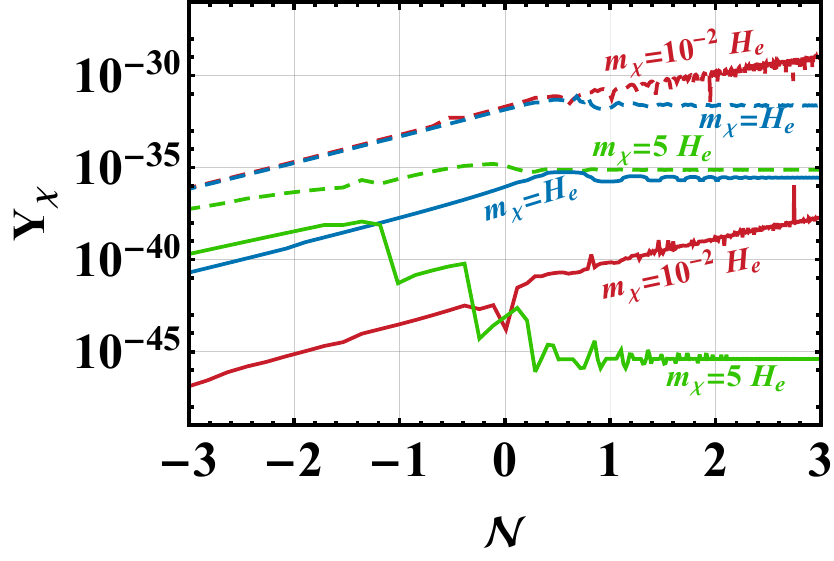}
\includegraphics[width=.32 \textwidth]{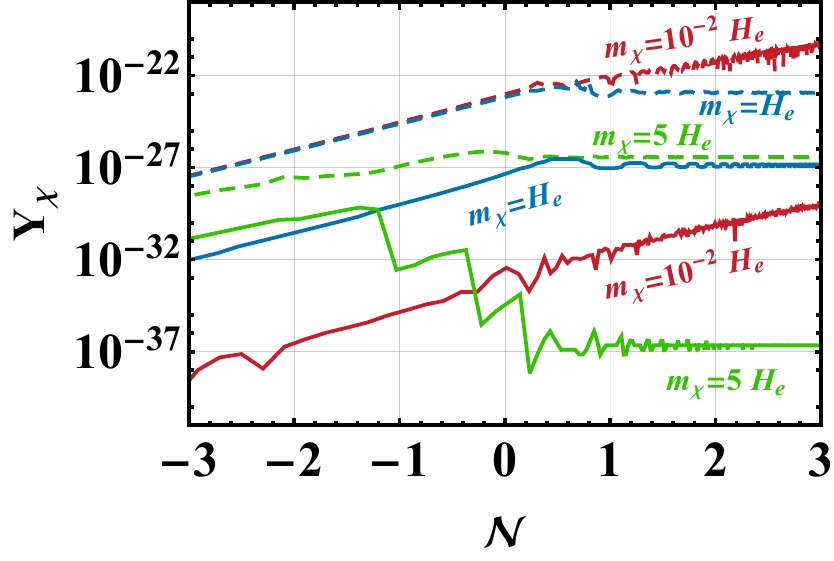}
\includegraphics[width=.32 \textwidth]{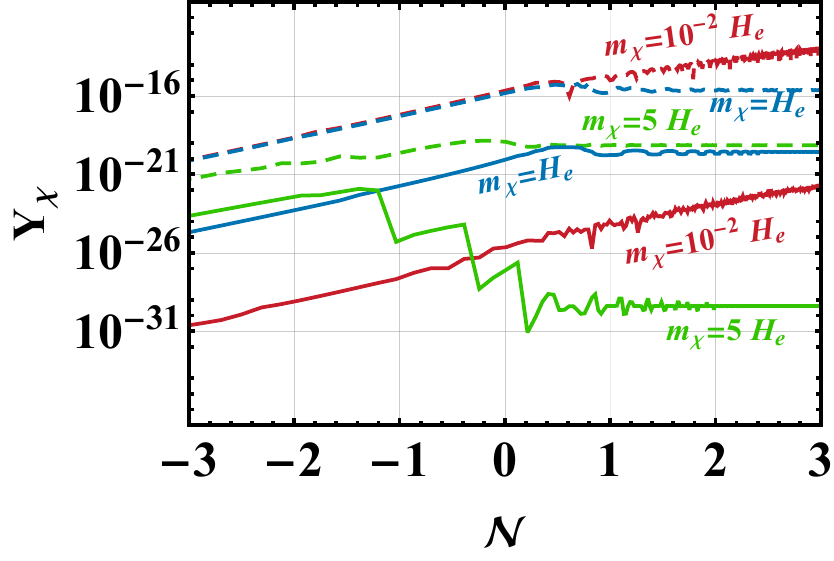}
  \caption{Evolution of the comoving number density of DM for the Benchmark points BP$a$,  BP$b$ and BP$c$ from left to right for $m_\chi= 10^{-2} \ H_e$ (red), $H_e$ (blue) and $5H_e$ (green). The dashed and solid lines correspond to $\xi_\chi=0$ and $\frac{1}{6}$, respectively.}
  \label{cmnumber}
\end{figure}

To determine the DM relic abundance, we consider the three benchmark points (BPs) listed in Table~\ref{table:bp}, fixing $\lambda=10^{-2}$ throughout the paper~\footnote{We have verified that the parameters in Table~\ref{table:bp} yield $A_s$, $n_s$, and $r$ consistent with observations~\cite{AtacamaCosmologyTelescope:2025blo,SPT-3G:2025bzu}. For their calculation in the
$R^2$-Higgs model, see Refs.~\cite{Modak:2025bjv,Lee:2021rzy}.}.  All BPs exhibit nearly single-field dynamics, with BP$a$ and BP$b$ lying in the mixed $R^2$-Higgs regime ($\xi_H \gg 1$), while BP$c$ lies in the $R^2$-like regime. Figure~\ref{cmnumber} shows the evolution of the comoving number density $Y_\chi$ with the $e$-folding number $\mathcal{N}$ for $m_\chi=10^{-2}H_e$ (red), $H_e$ (blue), and $5H_e$ (green), with $\xi_\chi=0$ (dashed) and $\frac{1}{6}$ (solid). For both couplings, $Y_\chi$ approaches an asymptotic value for $m_\chi=H_e$ and $5H_e$, whereas it continues to grow parametrically for $m_\chi=10^{-2}H_e$. This difference can be understood from the contributions to the effective mass squared, $m_{\mathrm{eff},(\chi)}^2$, during and after inflation.

\begin{figure}[h]
  \centering
\captionsetup{justification=raggedright,singlelinecheck=false}
\includegraphics[width=.45 \textwidth]{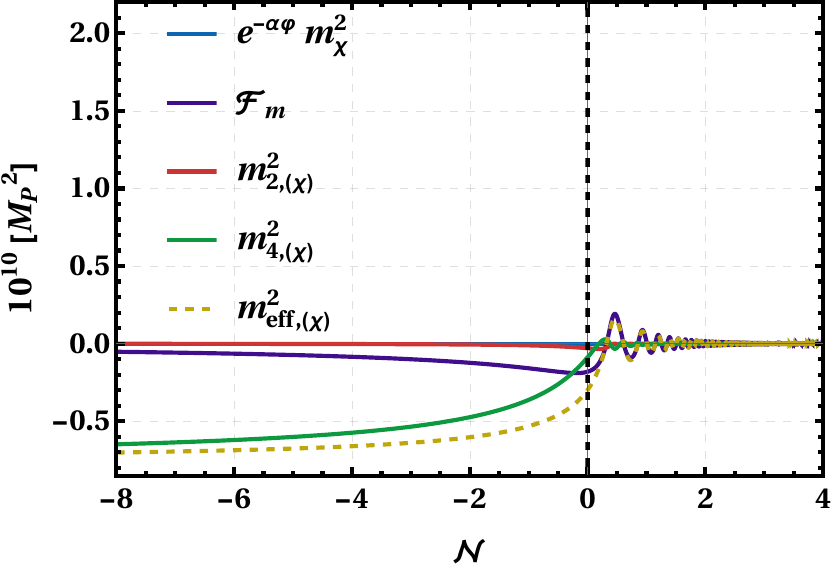}
\includegraphics[width=.45 \textwidth]{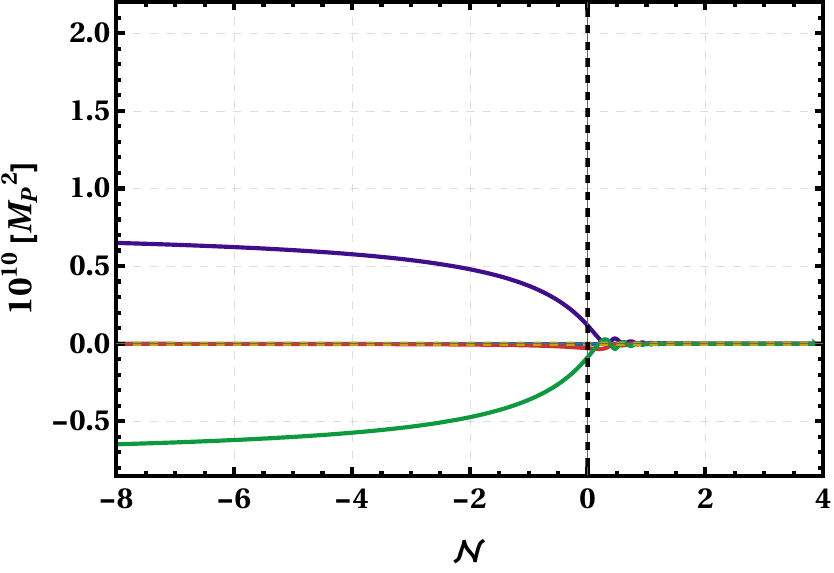}\\
\includegraphics[width=.45 \textwidth]{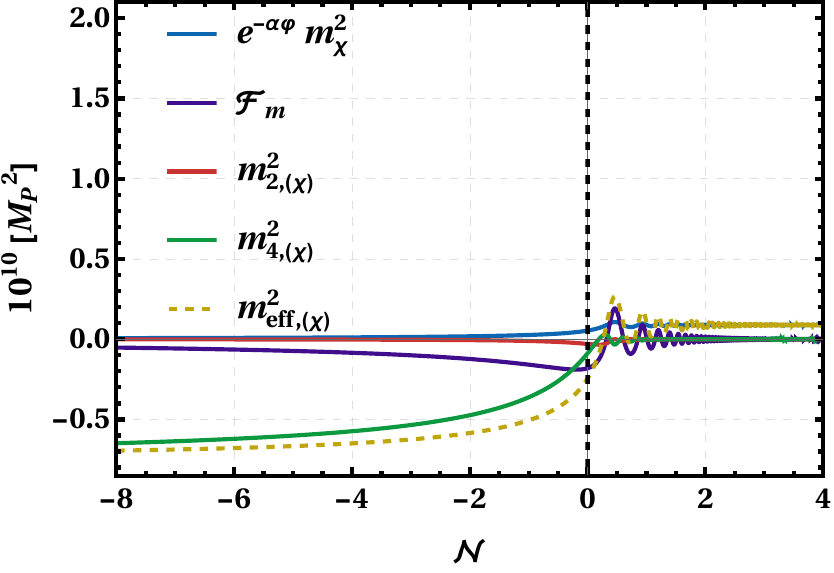}
\includegraphics[width=.45 \textwidth]{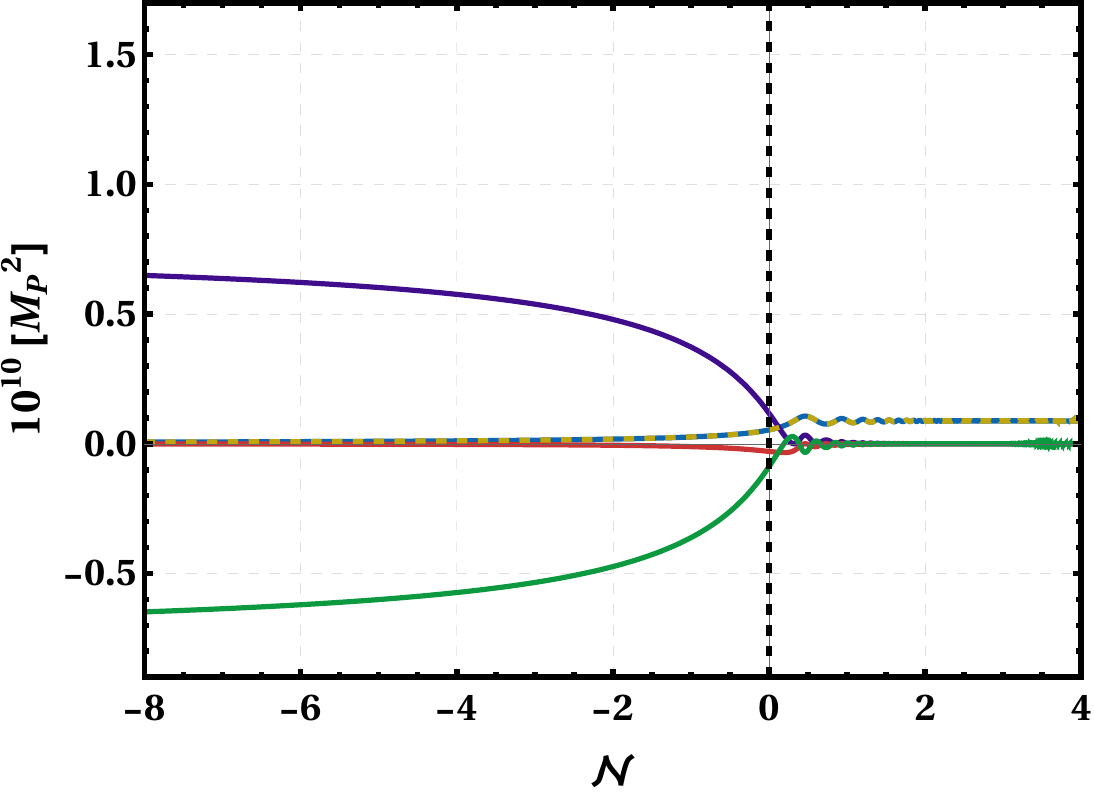}\\
\includegraphics[width=.45 \textwidth]{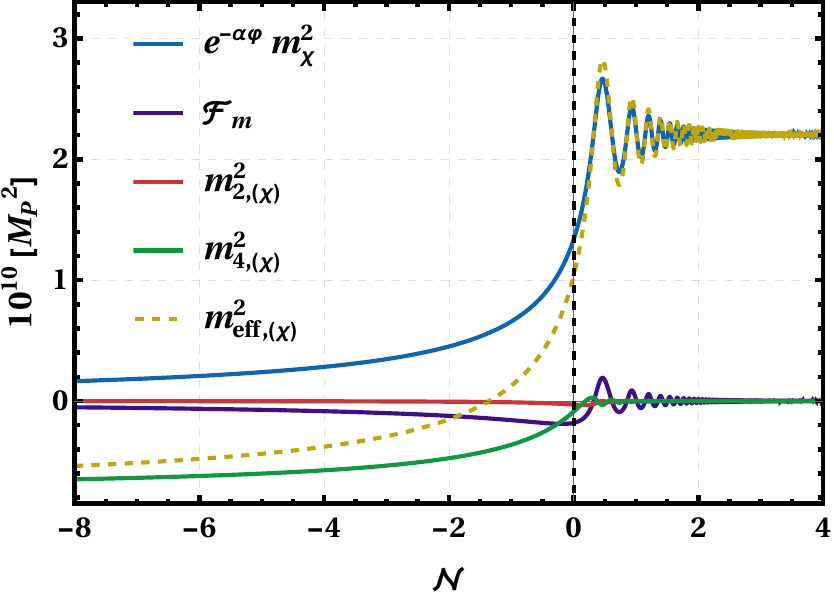}
\includegraphics[width=.45 \textwidth]{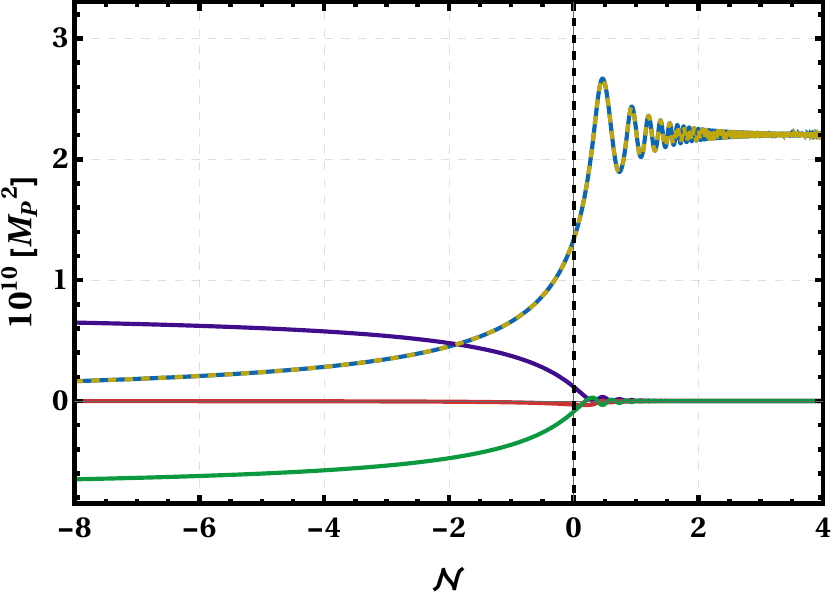}
  \caption{The effective mass $m_{\rm {eff},(\chi)}^2$ and its different contributions for BP$b$ for $\xi_\chi=0$ (left) and $\frac{1}{6}$ (right) with $m_\chi=0.01 \ H_e$ (upper), $H_e$ (middle) and $5H_e$ (lower) respectively. The different solid lines correspond to different contributions of $m_{\rm {eff},(\chi)}$. The vertical dashed line corresponds to the end of inflation.
}\label{massplot}
\end{figure}

To illustrate this behavior, Fig.~\ref{massplot} shows $m_{\mathrm{eff},(\chi)}^2$ and its individual contributions for BP$b$. For $\xi_\chi=0$ and $m_\chi=10^{-2}H_e$, the negative contribution $m_{4,(\chi)}^2$ dominates during inflation, keeping $m_{\mathrm{eff},(\chi)}^2$ tachyonic, as indicated by the olive dashed curve. Consequently, the effective frequency squared, $\omega_{(\chi)}^2$, becomes negative for modes that satisfy $\frac{k^2}{a^2} < |m_{\mathrm{eff},(\chi)}^2|$ with $m_{\mathrm{eff},(\chi)}^2<0$, leading to tachyonic growth of the mode function $z_k$. After the end of inflation, $\mathcal{F}_m$ dominates $m_{\mathrm{eff},(\chi)}^2$ and oscillates around zero because of its dependence on the oscillating background condensate $\varphi$; see the purple solid curve in the upper-left panel of Fig.~\ref{massplot}. As a result, $m_{\mathrm{eff},(\chi)}^2$ oscillates around zero, the frequency squared,
$\omega_{(\chi)}^2$, also oscillates, but becomes negative only when the negative mass contribution overcomes the momentum term. During these intervals, modes satisfying $k^2/a^2<|m_{\mathrm{eff},(\chi)}^2|$ undergo tachyonic amplification. However, during positive-frequency-squared intervals, the modes oscillate. Repeated variations produce parametric resonance in suitable momentum bands, even when $\omega_{(\chi)}^2$ remains positive. The effective mass squared therefore also oscillates around zero, driving continued parametric growth of $Y_\chi$. A similar behavior occurs for $\xi_\chi=\frac{1}{6}$ at the same mass, with $Y_\chi$ failing to reach an asymptotic plateau. This continued growth is also found for the other two BPs, as shown by the red solid and dashed curves in Fig.~\ref{cmnumber}.

For $m_\chi=H_e$ and $5H_e$, and for both $\xi_\chi=0$ and $\frac{1}{6}$, the negative contribution $m_{4,(\chi)}^2$ again dominates $m_{\mathrm{eff},(\chi)}^2$ during inflation, leading to tachyonic growth. For $m_\chi=H_e$ and $5H_e$, after the end of inflation $m_{\mathrm{eff},(\chi)}^2$ practically becomes constant as the term $e^{-\alpha\varphi}m_\chi^2$ dominates
keeping $m_{\mathrm{eff},(\chi)}^2$ positive and approximately constant by $\mathcal{N}\sim1$ (see middle and lower row of Fig.~\ref{massplot}).
As $m_{\mathrm{eff},(\chi)}^2\approx e^{-\alpha\varphi}m_\chi^2$ is positive and practically constant,
$\omega_{(\chi)}^2\approx k^2+a^2m_\chi^2$ still varies because the Universe expands. Therefore, a constant effective mass does not by itself imply that particle production stops.
As the $e^{-\alpha\varphi}\simeq 1$ and $\left|\frac{d m_{\mathrm{eff},(\chi)}^2}{dt}\right|
\ll 2H m_\chi^2$ after around $\mathcal{N}\gtrsim 1$, the adiabaticity parameter becomes
\begin{align}
\left|\frac{\omega_{(\chi)}'}{\omega_{(\chi)}^2}\right|
\approx\frac{H m_\chi^2}{\left(k^2/a^2+m_\chi^2\right)^{3/2}}.
\end{align}
For nonrelativistic modes, $k/a\ll m_\chi $, this reduces to $H/m_\chi$. Once $H < m_\chi$, evolution becomes practically adiabatic, further particle production is suppressed, and the comoving number density approaches a plateau. A similar feature is found for the other two BPs. This is reflected in $Y_\chi$, which reaches an asymptotic plateau for $\mathcal{N}\gtrsim1$ for both $m_\chi=H_e$ and $5H_e$, as shown in Fig.~\ref{cmnumber}.

These results indicate that dominance of $e^{-\alpha\varphi}m_\chi^2$ allows the comoving DM number density to stabilize within $\mathcal{N}\sim1$, a condition required for the validity of our treatment, as discussed below. For the $R^2$-like and mixed $R^2$-Higgs benchmarks in Table~\ref{table:bp}, this occurs for $m_\chi\gtrsim H_e$ at both $\xi_\chi=0$ and $\frac{1}{6}$. Consistently, for all three BPs, no region with $m_\chi< H_e$ in which $Y_\chi$ reaches an asymptotic plateau for either coupling $\xi_\chi=0$ or $\xi_\chi=\frac{1}{6}$ is found within $\mathcal{N}\lesssim 3$ (see Fig.~\ref{cmnumber}).

\begin{figure}[h]
  \centering
\captionsetup{justification=raggedright,singlelinecheck=false}
\includegraphics[width=.45 \textwidth]{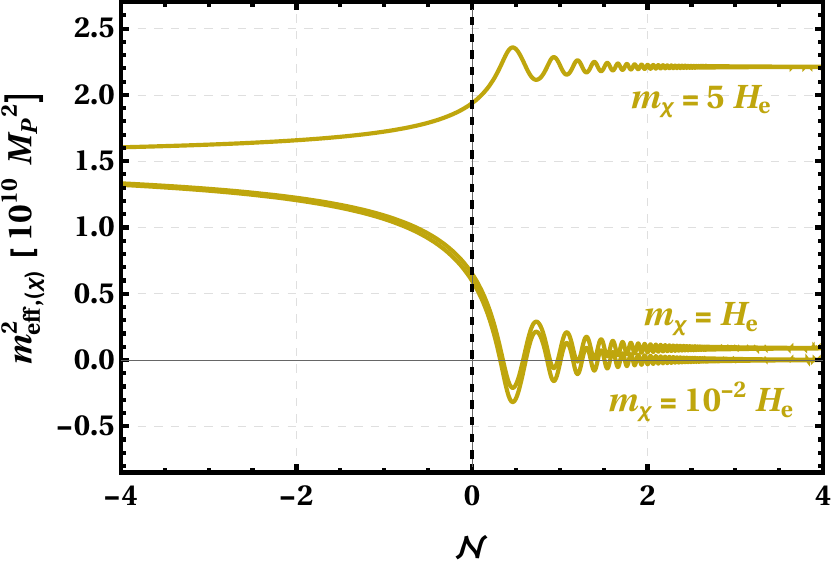}
\includegraphics[width=.45 \textwidth]{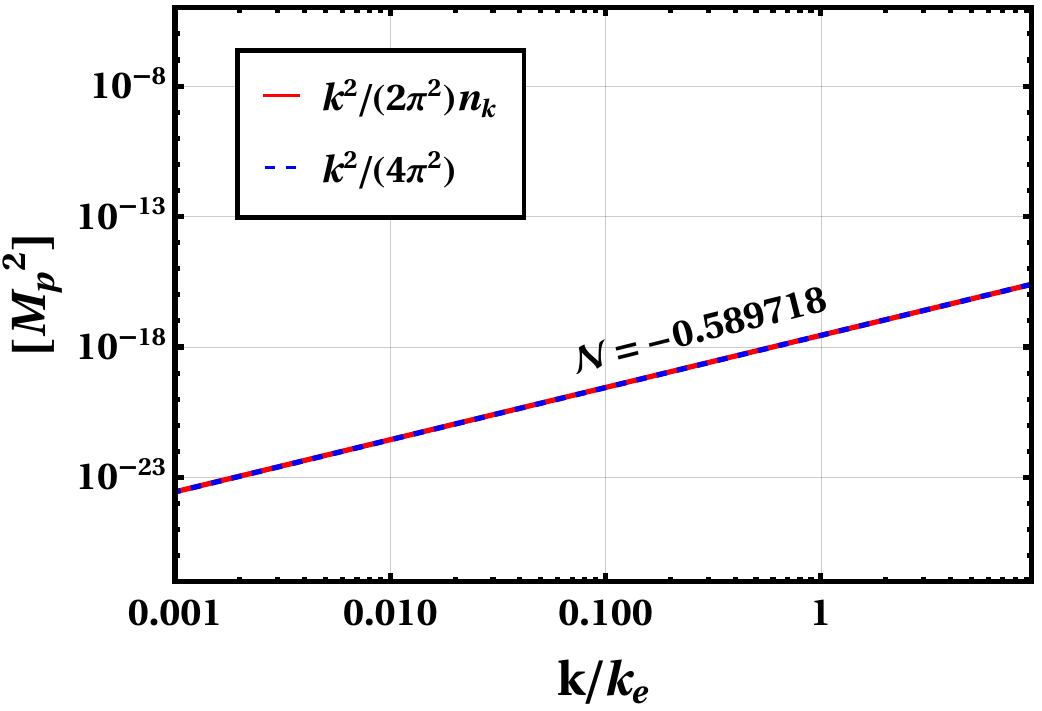}
  \caption{The effective mass $m_{\rm {eff},(\chi)}^2$ (left) the spectrum i.e. the integrand of Eq.~\eqref{eq:ychi} (right) for BP$b$ with $\xi_\chi = 1/2$. See text for more details.
}\label{plotxichi1by2}
\end{figure}

For $\xi_\chi>\frac{1}{6}$, we find no appreciable DM production during or after inflation for any BP. Figure~\ref{plotxichi1by2} illustrates this for BP$b$ with $\xi_\chi=1/2$, showing $m_{\mathrm{eff},(\chi)}^2$ and the spectrum, defined as the integrand of Eq.~\eqref{eq:ychi}. For $m_\chi=10^{-2}H_e$, $H_e$, and $5H_e$, the effective mass squared remains positive and nearly constant during inflation, oscillates mildly afterward, and reaches a plateau for $\mathcal{N}\gtrsim1$. The right panel shows that the mode spectrum (red solid) coincides with the BD vacuum spectrum (blue dashed) at $\mathcal{N}=-0.59$ for BP$b$ and $m_\chi=H_e$ scenario, indicating negligible excitation. We find similar behavior for the other two mass regime, with no appreciable excitation after inflation either. This is due to the positivity nature of $m_{\rm {eff},(\chi)}^2$ during inflation,  prevents tachyonic growth, while a sufficiently large effective mass suppresses excitation from expansion. After inflation, mild oscillations also preserve adiabaticity as the amplitude is sufficiently small; as a result no particle production occurs. Same feature is found for the other two BPs for $\xi_\chi = 1/2$ and for $m_\chi=10^{-2}H_e$, $m_\chi=H_e$ and  $m_\chi=5H_e$.

\subsection{Dark Matter Relic Abundance}
The gravitational production of the spectator field during and after inflation can generate a population of DM particles even in the absence of direct interactions between the spectator field and the visible sector. Once the production of $\chi$ particles becomes negligible and the comoving number density freezes, the subsequent evolution of the relic abundance is determined by the expansion of the Universe. In this section, we calculate the present-day relic abundance resulting from the gravitational production mechanism and investigate its dependence on the spectator-field mass $m_\chi$ and the non-minimal coupling $\xi_\chi$.

The present-day energy density can be expressed in terms of the physical number density as
\begin{equation}
\rho_{\chi,0}=m_\chi n_{\chi,0}\,.
\end{equation}
Since the spectator particles are non-relativistic at late times, their energy density is well approximated by $m_\chi n_\chi$. It is convenient to work with the comoving number density,
$Y_\chi$,
which remains approximately constant after the particle production process has ceased. Therefore, the present-day physical number density can be obtained from the frozen comoving number density as
\begin{equation}
n_{\chi,0}=\frac{Y_\chi^{\rm f}}{a_0^3},
\end{equation}
where $Y_\chi^{\rm f}$ denotes the asymptotic value of the comoving number density after gravitational particle production has become inefficient. Therefore, the present-day DM relic abundance is given by,
\begin{equation}{\label{omegachi1}}
    \Omega_\chi h^2=\frac{\rho_{\chi,0}}{\rho_c/h^2}=\frac{m_\chi\,Y_\chi^{\rm f}}{\rho_c/h^2}\frac{1}{a_0^3}
\end{equation}
where $\rho_c/h^2=8.1\times10^{-47}\,\mbox{GeV}^4$ \cite{ParticleDataGroup:2026mpi} and $a_0$ are the present-day critical energy density and scale-factor, respectively. It is also useful to express the abundance in terms of the preheating temperature $\tpre$. We have assumed here that thermalization is complete after the end of preheating. Using the fact that from the end of preheating to the present time, the total comoving entropy density of the universe is conserved, which  leads to~\cite{Dai:2014jja,Cook:2015vqa}
\begin{equation}\label{acmb}
    a_0=\apre\,\left(\frac{g_{*s}(\tpre)}{g_{\star s}(T_0)}\right)^{1/3}\,\frac{\tpre}{\t0}
\end{equation}
where $g_{*s}(\tpre)=106.75$ and $g_{*s}(\t0)\simeq3.91$ are the effective number entropic degrees of freedom at the time of preheating and the present time, respectively. $\t0$ be the present-day CMB temperature, which is $2.35\times10^{-13}$ GeV. Using the relation~\eqref{acmb} in Eq.~\eqref{omegachi1}, the relic abundance can be rewritten as,
\begin{equation}\label{eq:abundance}
\omegax=5.86\times10^6\,\left(\frac{\mx}{\rm GeV}\right)\,\frac{Y_\chi^{\rm f}}{(\apre\,\tpre)^3}
\end{equation}
\begin{figure}[h]
  \centering
\captionsetup{justification=raggedright,singlelinecheck=false}
     \includegraphics[width=.5 \textwidth]{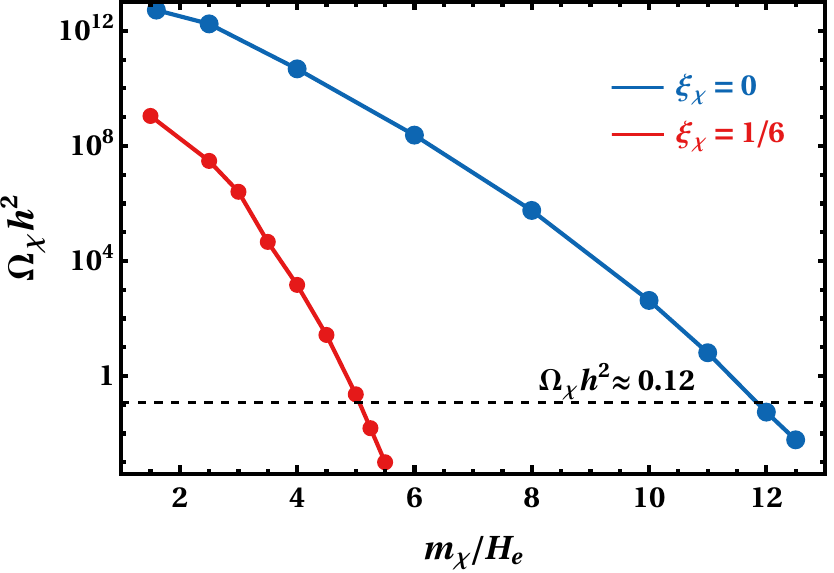}
      \caption{The relic abundance as a function of spectator mass $\mx$ for BP$b$ for the value of $\xi_\chi=\frac{1}{6}$(red) and $\xi_\chi=0$(blue). See text for details.}
  \label{relicplot}
   \end{figure}

Thus, once the asymptotic comoving number density generated by gravitational particle production and the $\tpre$ is found, the corresponding present-day DM abundance can be calculated directly from Eq.~\eqref{eq:abundance}. We remark that many studies in the literature take the $\tpre$ (or reheating) temperature as a free parameter in $R^2$-Higgs model. We, however, show that for the parameter space discussed in Table~\ref{table:bp}, the $h$ and Goldstones $\phi_2$, $\phi_3$ and $\phi_4$ can have successful preheating within $\mathcal{N}\lesssim 2$ $e$-folding after the end of inflation with $\tpre\simeq\mathcal{O}(10^{14})$ GeV. We remark that, in the $R^2$-like regime i.e. for $\xi_R\gg\xi_H$, such preheating is induced as a by product if $\xi_H\gtrsim 1$. In addition, as the linear theory breaks down after the completion of preheating, the asymptotic value of $Y_\chi$ (i.e. $Y_\chi^{\rm f}$) should be reached before the completion of preheating.  For the super-Hubble $\mx$ and $\mx \sim H_e$, this asymptotic behavior of $Y_\chi$, can be seen in Fig.~\ref{cmnumber}, where $Y_\chi$ reaches its asymptotic value $\mathcal{N}\simeq 1.2$ $e$-folding after the end of inflation. As we shall see shortly, this is well before the completion of preheating for all the BPs in Table~\ref{table:bp}.

Preheating thus plays two key roles: (i) for a given particle content, it fixes the preheating temperature and hence $\Omega_\chi h^2$; (ii) it sets the time frame within which the comoving number density should reach an asymptotic plateau, as required for the validity of linear theory. This emphasises that the parameter space for gravitational DM production from a spectator field is tightly constrained if we do not take $\tpre$ as a free parameter and require the validity of linear-order theory. Our findings rely on the linear-order approximation and may change when backreaction, rescattering, and decay are included. Further, for $\xi_H\ll1$, preheating through Higgs and Goldstone modes is incomplete, and reheating proceeds perturbatively, with a temperature still fixed by the model. We leave the impact of perturbative reheating regime on gravitational DM production from a spectator field for future work, as it requires treating condensate decay and nonlinear, non-equilibrium effects. For $\xi_H\gtrsim1$, however, preheating may play a crucial role in gravitational DM production. This connection remains largely unexplored in the $R^2$-Higgs model and, to the best of our finding, many other inflationary models.

Figure~\ref{relicplot} shows the present-day DM relic abundance as a function of $m_\chi$ for BP$b$, with $\xi_\chi=\frac{1}{6}$ (red) and $\xi_\chi=0$ (blue) as representative examples. In order to find Fig.~\ref{relicplot}, we numerically solve the mode equation Eq.~\eqref{dmmode} for the spectator DM field utilizing the Bunch-Davis (BD) initial condition, and obtain the corresponding occupation number and comoving number density. We then follow the evolution of $Y_\chi$ until it reaches its asymptotic value after the end of inflation and finally estimate the relic density utilizing Eq.~\eqref{eq:abundance}. While generating Fig.~\ref{relicplot}, we have assumed preemtively $\tpre
\sim5\times10^{14}$ GeV. We shall see shortly in the following section that this is indeed the ballpark preheating temperature BP$b$. For both $\xi_\chi=\frac{1}{6}$ and 0, we find that the present-day relic abundance decreases monotonically with increasing $m_\chi$ in the super-Hubble $\mx$. This behavior can be understood from the exponential suppression of gravitational particle production for increasingly heavy spectator, which approximately scales as $e^{-2\pi m_\chi/H_e}$ \cite{Li:2019ves,Ema:2018ucl,Verner:2024agh}. Consequently, there is a maximum spectator mass beyond which gravitational production becomes insufficient to account for the observed relic abundance. For $\xi_\chi=\frac{1}{6}$ and $\xi_\chi=0$, we find the corresponding maximum masses to be $m_\chi\simeq5.05\ H_e=3.5\times10^{13}$ GeV and $m_\chi\simeq11.8H_e=8.4\times10^{13}$ GeV, respectively, for which the gravitationally produced spectator can account for the observed relic abundance.

Finally, we comment briefly on the production of sub-Hubble mass spectator particles, $\mx<\he$. Our analysis is performed at linear order, and this treatment is expected to become unreliable once nonlinear preheating becomes significant. Therefore, the gravitational production of $\chi$ should reach an asymptotic value before the completion of preheating. In the sub-Hubble mass regime, however, particle production continues long after preheating, since the frequency mode oscillates across zero with a long period after preheating, as discussed above. Moreover, in this regime, the effective mass is predominantly determined by the background-induced contributions such as $\mathcal{F}_m$ and shows little sensitivity to the bare mass $\mx$. Of course, there may exist fine cancellation of different terms within $m_{\mathrm{eff},(\chi)}^2$, which may make $\mx$ dependence dominant and provide asymptotic $Y_\chi$ value within $\mathcal{N}\simeq 1.2$ $e$-folding, however in our numerical scan, we have not identified a such parameter space in which the bare mass term controls the effective mass after inflation. We leave this investigation for future work.

\section{Higgs and Goldstone Preheating}
\label{sec:prehiggs}

Preheating, the early stage of reheating, can lead to nonperturbative particle production, thereby modifying the thermal history and subsequent evolution of the Universe. The mechanism is analogous to gravitational DM production, except that the particles are produced by the rapidly oscillating background condensates. The relic abundance depends on $\tpre$ and the duration of preheating. While $\tpre$ directly enters Eq.~\eqref{eq:abundance}, the duration determines whether DM production remains within the validity of the linear-order approximation. This applies to both DM and other particles, including those of the SM. As discussed in the preceding section, we assumed $\tpre\sim 5\times10^{14}$ GeV to calculate the relic abundance shown in Fig.~\ref{relicplot}. We now investigate whether preheating can lead required $\tpre$ not only for BP$b$ but also for the other two BPs. In baseline $R^2$-Higgs inflation, successful preheating within $1.5$-$2$ $e$-folding is possible from $h$, $\phi_2$, $\phi_3$, and $\phi_4$, with $\tpre\sim10^{14}$ GeV, for similar parameter values~\cite{Cado:2024von}. Whether in the extended model such as ours yields a comparable $\tpre$  and preheating duration remains to be seen, particularly to ensure that $Y_\chi$ freezes out before the linear-order approximation breaks down.

We first proceed towards quantization of $h$. Due to the presence of nonminimal coupling $\xi_H$, the EoMs of the Higgs and the inflaton are coupled with each other, as can be seen from
$\mathcal{M}^I_{~~J}$ term in Eq.~\eqref{eq:perturb}. Therefore, the quantization of $\phi$ and $h$ needs some additional care. We quantize the fluctuations $\hat{\widetilde{X}}^{\phi}(\tau,\vb{k})$ and $\hat{\widetilde{X}}^{h}(\tau,\vb{k})$ by decomposing in momentum space as~\cite{Amin:2014eta,DeCross:2015uza,Sfakianakis:2018lzf,Cado:2024von}
\begin{align}
\hat{\widetilde{X}}^I(\tau, \vb{k})=  \sum_m \left[ u_m^I(\tau,k) \hat{a}_m(\vb{k}) + u_m^{I*}(\tau,k) \hat{a}^\dagger_m(-\vb{k})   \right],\label{eq:Xmuquantization}
\end{align}
where $m \in \{1,2\}$ denotes  $\{\phi,h\}$. We parametrize the mode function           6s $u_m^I(\tau,k)$ as
\begin{align}
u_m^I(\tau,k) = t_{(m,I)}(\tau,k) e_m^I(\tau),
\end{align}
where $e_m^I(\tau)$ are the vielbeins of the field-space metric and the $t_{(m,I)}(\tau,k)$ are in general complex scalar functions.
We take  $t_{(1,\phi)}(\tau,k)=v_{1k}(\tau)$, $t_{(2,\phi)}(\tau,k)=v_{2k}(\tau)$, $t_{(1,h)}(\tau,k)=y_{1k}(\tau)$
and $t_{(2,h)}(\tau,k)=y_{2k}(\tau)$. The annihilation and creation operators $\hat{a}_m(\vb{k})$ and $\hat{a}^\dagger_m(-\vb{k})$ satisfy the usual commutator relationships
with
\begin{align}
 \hat{a}_m(\vb{k})  \ket{0} = 0,\hspace{3 cm} \bra{0} \hat{a}^\dagger_m(\vb{k}) = 0.
\end{align}

The BPs in Table~\ref{table:bp} resemble single-field regime, where the off-diagonal elements $\mathcal{M}^\phi_{~~h}$ and $\mathcal{M}^h_{~~\phi}$ are much smaller than the diagonal elements $\mathcal{M}^\phi_{\phi}$ and $\mathcal{M}^h_{h}$. Consequently, the vielbeins are approximately diagonal and $\widetilde{X}^\phi$ and $\widetilde{X}^h$ depend only on $v_{1k}(\tau)$ and $y_{2k}(\tau)$ respectively. The mode equations therefore reduce to~\cite{Cado:2024von}
\begin{align}
&v_{1k}'' + \omega^2_{(\phi)}\,v_{1k} \simeq 0,\label{eq:inflatonfluc}\\
&y_{2k}'' + \omega^2_{(h)}\,y_{2k} \simeq 0,\label{eq:higgsfluc}
\end{align}
where $\omega^2_{(I)}=k^2 + a^2 m_{\mathrm{eff},(I)}^2$ with $I = \phi, h$ and, $m_{\mathrm{eff},(I)}^2=\sum_i m_{i,(I)}^2$ as in before.
The energy densities per mode for $\phi$ and $h$ fluctuations are
\begin{subequations} \begin{eqnarray}
\rho_k^{(\phi)} &=& \dfrac{1}{2} G_{\phi\phi}\left(|v'_{1k}|^2 +\omega^2_{(\phi)}   |v_{1k}|^2\right) e_1^\phi e_1^\phi = \dfrac{1}{2}\left(|v'_{1k}|^2 +\omega^2_{(\phi)}   |v_{1k}|^2\right) , \label{enphi_conf-simp}\\
 \rho_k^{(h)} &=& \dfrac{1}{2} G_{hh}\left(|y'_{2k}|^2 +\omega^2_{(h)}  |y_{2k}|^2 \right)e_2^h e_2^h = \dfrac{1}{2}\left(|y'_{2k}|^2 +\omega^2_{(h)}  |y_{2k}|^2 \right), \label{enh_conf-simp}\\
 \rho_k^{\mathrm{int}}& =& \mathcal{O}(h^2)\sim 0.  \label{en-phi+h_conf-simp}
\end{eqnarray}  \end{subequations}
The vacuum subtracted physical quantum energy densities $\rho^q_{(\phi)}$ and $\rho^q_{(h)}$
\begin{align}
\rho^q_{(I)} = \frac{1}{a^4}\int\left( \frac{d^3k}{(2\pi)^3} \rho_k^{(I)} - \frac{k^3}{4\pi^2 } dk \right), \label{eq:quantumphih}
\end{align}

In order to compute $\rho^q_{(h)}$ numerically, we solve Eq.~\eqref{eq:higgsfluc} utilizing the BD initial condition for all subhorizon modes at least three $e$-folds before the end of inflation. We include only modes within the horizon at each time, as superhorizon modes remain frozen and do not contribute to thermalization~\cite{Sfakianakis:2018lzf,Cado:2024von}. We redirect readers for the details of the adaptive numerical framework to Ref.~\cite{Cado:2024von}.

\begin{figure}[h]
  \centering
\captionsetup{justification=raggedright,singlelinecheck=false}
\includegraphics[width=.45 \textwidth]{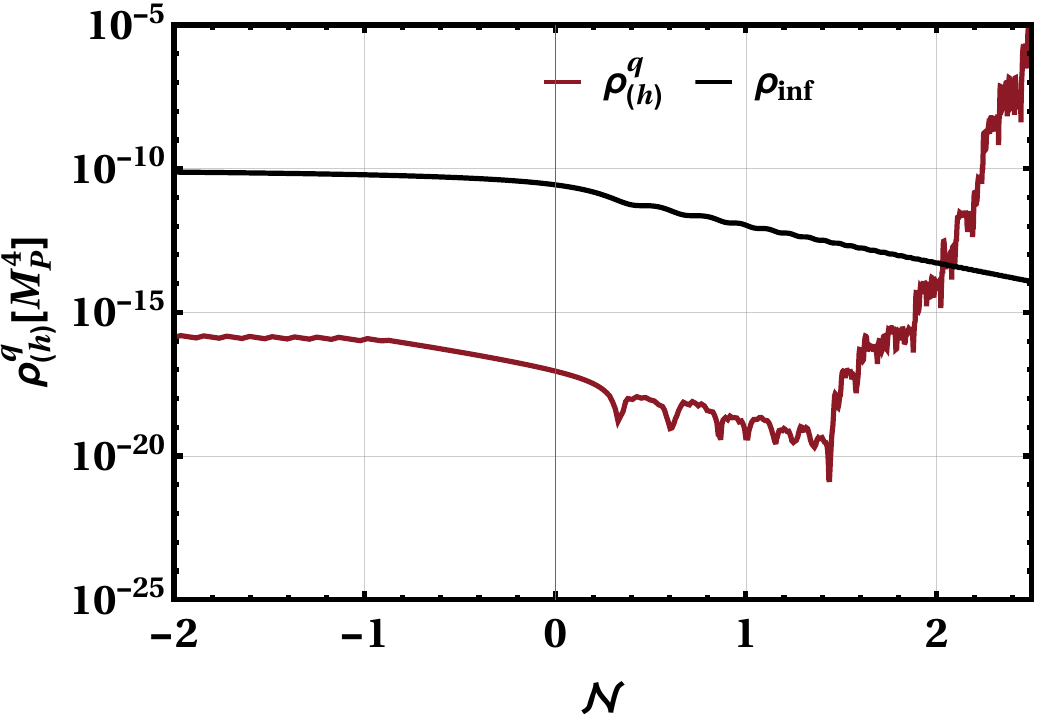}
\includegraphics[width=.45 \textwidth]{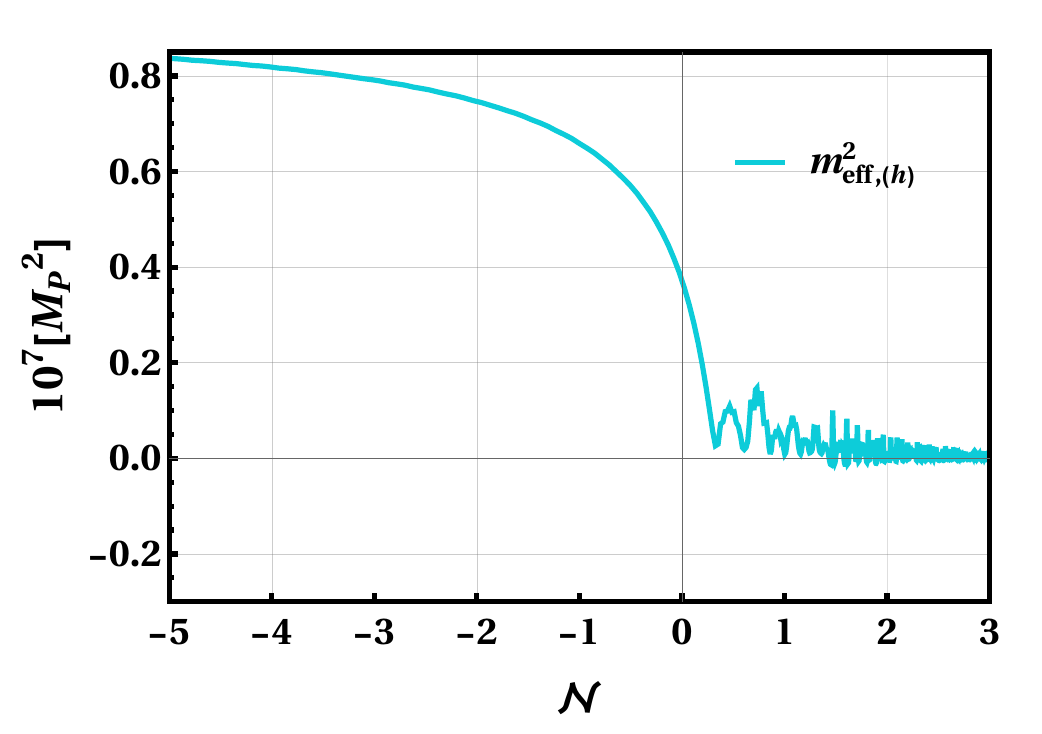}\\
\includegraphics[width=.45 \textwidth]{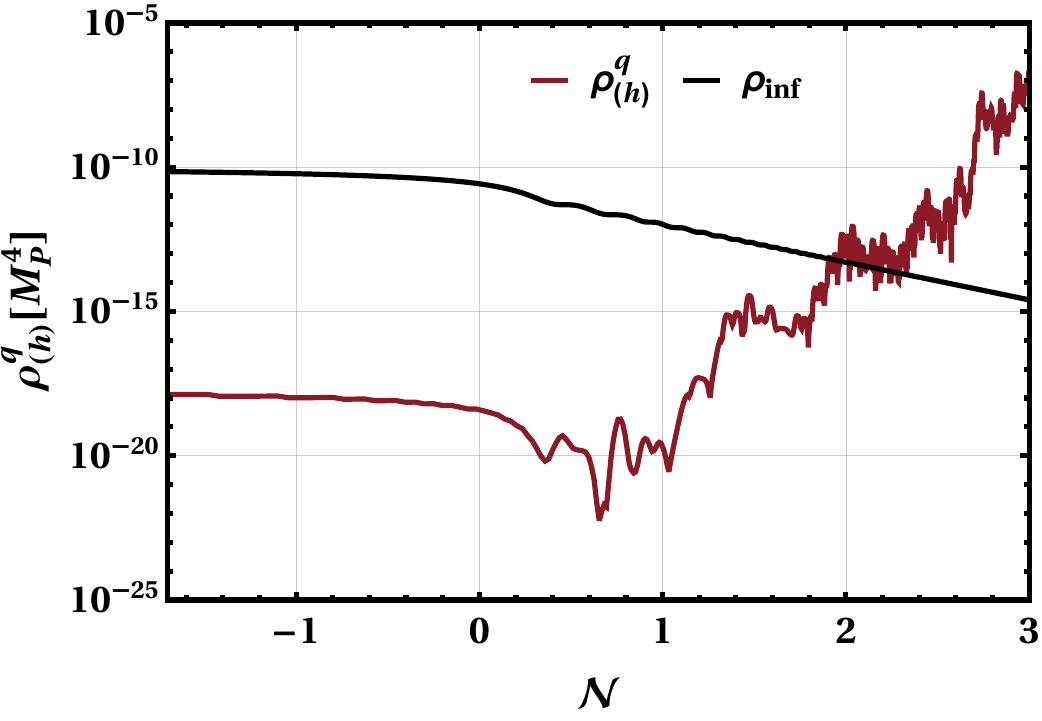}
\includegraphics[width=.45 \textwidth]{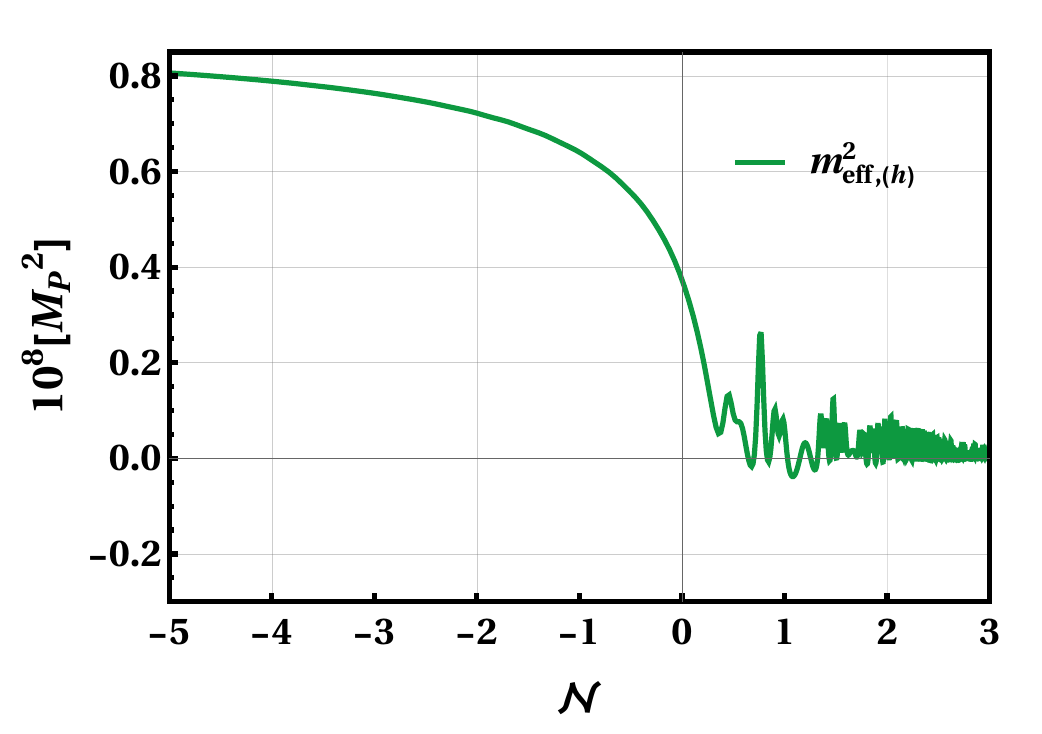}
\caption{The vacuum subtracted $\rho_{(h)}^{q}$ and $m_{\mathrm{eff},(h)}^2$ respectively for BP$a$ (upper panel) and BP$b$ (lower panel) respectively. See text for more details. The background energy densities $\rho_{\rm inf}$  are plotted for both the BPs in black.}
    \label{plot:higgs}
\end{figure}

In Fig.~\ref{plot:higgs}, we plot $\rho^q_{(h)}$ for BP$a$ and BP$b$, along with the corresponding $\rho_{\mathrm{inf}}$ from Eq.~\eqref{eq:infenergy}. For both BPs, successful preheating, defined by $\rho^q_{(h)}\simeq\rho_{\rm inf}$, occurs within $\mathcal{N}=1.8$-$2$ $e$-folds after inflation. This behavior arises from parametric resonance induced by the oscillations of $m_{\mathrm{eff},(h)}^2$ as $h_0$ oscillates around zero. Although $m_{\mathrm{eff},(h)}^2$ becomes negative for $\mathcal{N}\gtrsim 1.5$, its amplitude is too small to trigger significant tachyonic growth.

We find that preheating from $h$ is not possible for BP$c$. Furthermore, $\phi$ fails to induce successful preheating for any of the three BPs, which is consistent with Ref.~\cite{Cado:2024von}. Although Higgs preheating is not possible, the Goldstone modes may lead to successful preheating for $\xi_H\sim 1$, as found in the baseline $R^2$-Higgs inflation model~\cite{Cado:2024von}. To examine whether this remains true in our scenario, we now focus on the EoMs of $\phi_2$, $\phi_3$, and $\phi_4$ as given in Eq.~\eqref{eq:perturb}. We restrict our discussion to the Goldstone mode $\phi_2$, since $\phi_3$ and $\phi_4$ yield similar results. We first proceed with the quantization of $\phi_2$, but there is a subtlety in the choice of gauge.

The unitary gauge, commonly used in the $R^2$-Higgs model, becomes ill-defined at every zero crossing of the background field $h_0$ after inflation~\cite{Sfakianakis:2018lzf,Cado:2024von}. This can be seen from Eq.~\eqref{eq:perturb} and the explicit expression for $\mathcal{A}_{(\phi_2)}$. A similar issue arises for $\phi_3$ and $\phi_4$. We therefore adopt the Coulomb gauge, $\partial_i Z^i=0$ and $\partial_i W^{\pm i}=0$, which remains well-defined at the zero crossings of $h_0$~\cite{Sfakianakis:2018lzf,Cado:2024von}. In this gauge, either the Goldstone fields or the longitudinal gauge bosons can be chosen as dynamical variables. In the following, we retain the Goldstone fields as the dynamical variables.
The decoupled doubly covariant rescaled perturbation as $X^{\phi_2}$ quantized in momentum space as
\begin{align}
\hat{\widetilde{X}}^{\phi_2} = s_k(\tau) e^{\phi_2}(\tau) \hat{a}(\vb{k}) + s^*_k(\tau) e^{\phi_2}(\tau) \hat{a}^\dagger(-\vb{k}), \label{eq:phi2quant}
\end{align}
where $s_k$ is the mode function,  $e^{\phi_2}$ corresponding vielbein and, $\hat{a}$ and $\hat{a}^\dagger$ are the annihilation and creation operators for the field $\phi_2$.

The EoM for mode $s_k$ is found from the Eq.~\eqref{eq:perturb} and the quantization Eq.~\eqref{eq:phi2quant} as~\cite{Sfakianakis:2018lzf,Cado:2024von}
\begin{align}
& s_k'' + \frac{ 2 m_Z^2 \Upsilon}{\mathcal{K}_Z} s_k' + \bigg[ k^2  + a^2 M_{\phi_2}^2+ \frac{ 2 m_Z^2 \Upsilon^2}{\mathcal{K}_Z}\bigg] s_k  = 0,\label{eq:goldmode}
\end{align}
where $M_{\phi_2}^2=m_{\mathrm{eff},(\phi_2)}^2+ m_Z^2$, $m_Z^2=  (g_Z^2/4)  e^{-\sqrt{\frac{2}{3}}\frac{\varphi}{M_{\rm P}}} h_0^2$ and,
\begin{align}
&\mathcal{K}_Z =\frac{k^2}{a^2}+ m_Z^2,~~\Upsilon (\tau) = \frac{\varphi'}{\sqrt{6}M_{\rm P}} -  \frac{a'}{a} -   \frac{h_0'}{h_0}.
\end{align}
We remark here that while finding Eq.~\eqref{eq:goldmode} we have rescaled $Z_0 \to Z_0/a$ and utilized the Coulomb gauge condition $\partial_i Z^i=0$.

The vacuum-subtracted quantum energy density for $\phi_2$ is given as~\cite{Cado:2024von}
\begin{align}
\rho^q_{(\phi_2)} = \frac{1}{a^4}\int\left( \frac{d^3k}{(2\pi)^3} \rho_k^{(\phi_2)} - \frac{k^3}{4\pi^2 \Delta_{(\phi_2)} } dk \right), \label{eq:quantumphi}
\end{align}
where
\begin{align}
\rho_k^{(\phi_2)}\; &= \;   \frac{1}{2} \Biggl\{ \left(1-\frac{m_Z^2}{\mathcal{K}_Z} \right)\left|s_k^\prime\right|^2
 +  \Biggl[k^2+a^2 m_{\mathrm{eff},(\phi_2)}^2- \frac{m_Z^2}{\mathcal{K}_Z} \;\Upsilon^2\Biggl] \left|s_k\right|^2
- \frac{m_Z^2}{\mathcal{K}_Z} \;\Upsilon \; (s'_k s^\ast_k +s^{\ast\prime}_k s_k )
\Biggr\},\label{eq:golphi2}
\end{align}
with
\begin{align}
\Delta_{(\phi_2)} = \exp{\int_{-\infty}^{\tau}\frac{ 2 m_Z^2 \Upsilon}{\mathcal{K}_Z}\,d\tau'}.
\end{align}

\begin{figure}[h]
  \centering
\captionsetup{justification=raggedright,singlelinecheck=false}
      \includegraphics[width=.45 \textwidth]{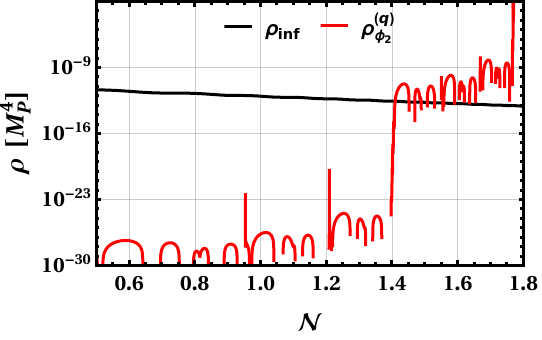}
    \includegraphics[width=.49 \textwidth]{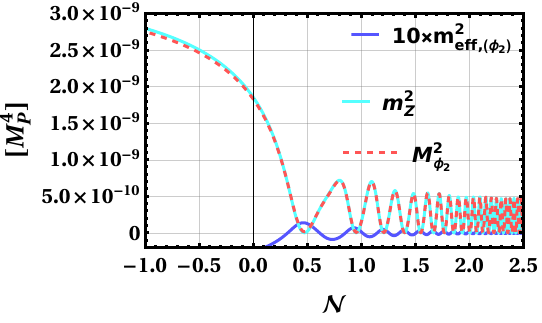}
  \caption{The $\rho^q_{(\phi_2)}$ (left panel) and effective mass $M_{\phi_2}^2$ (right panel) as a function of $\mathcal{N}$ for BP$c$.
The background energy density $\rho_{\rm inf}$ is plotted in black in the left panel. }
 \label{plot:energyG}
\end{figure}

In Fig.~\ref{plot:energyG}, we plot $\rho^q_{(\phi_2)}$ as a function of $\mathcal{N}$ along with $\rho_{\rm inf}$ for BP$a$. It is clear that $\phi_2$ can lead to successful preheating within $\mathcal{N}\sim 1.5$. Unlike the Higgs mode $h$, preheating through $\phi_2$ is primarily governed by $m_Z^2$, which provides the dominant contribution to $M_{\phi_2}^2$, as shown in the right panel of Fig.~\ref{plot:energyG}. Since $m_Z^2$ is directly proportional to $h_0^2$, the oscillations of $h_0$ lead to oscillations in $m_Z^2$, resulting in parametric amplification of $\rho^q_{(\phi_2)}$. The Goldstone modes $\phi_3$ and $\phi_4$ exhibit similar preheating behavior, and hence their results are not shown here.

\begin{table}[h]
\centering
\begin{tabular}{|c |c| c| c| c| }
    \hline
	BP  & preheating field    & $\mathcal{N}_{\rm pre}$ & $\rho_{\rm pre} \; [M_{\rm P}^4]$ &  $T_{\rm pre}$ [GeV]\\
     \hline
        $a$ & $h$  &  $2.1$ &  $5.94\times10^{-14}$ & $4.86\times 10^{14}$ \\
   \hline
        $b$ & $h$  &  $1.9$ &  $6.80\times10^{-14}$ & $5.03\times 10^{14}$ \\
        \hline
        $c$ &$\phi_2$  & $1.5$  & $7.8 \times 10^{-14}$  &  $7.2 \times 10^{14}$ \\
	\hline
	\end{tabular}
	\caption{The details of preheating for all three BPs chosen for our analysis.}
	\label{table:reheating}
\end{table}

Assuming that thermalization is completed immediately after preheating, we estimate the preheating temperature $T_{\rm pre}$ using
\begin{align}
\rho_{\mathrm{inf}}\bigr|_{\mathcal{N}=\mathcal{N}_{\rm pre}}
\equiv \rho_{\rm pre}
=\frac{g_{\rm pre}\pi^2}{30}\,T_{\rm pre}^4,
\label{eq:prehettemp}
\end{align}
where $\mathcal{N}_{\rm pre}$ is the number of $e$-folds required to complete preheating, and $g_{\rm pre}=106.75$ is the number of relativistic degrees of freedom at the completion of preheating. The corresponding values of $T_{\rm pre}$ and $\mathcal{N}_{\rm pre}$ for all three BPs are summarized in Table~\ref{table:reheating}. For BP$a$ and BP$b$, the Goldstone boson $\phi_2$ yields a preheating temperature and $\mathcal{N}_{\rm pre}$ comparable to those for BP$c$. For simplicity, we use $T_{\rm pre}$ from $h$ to estimate the DM relic abundance for these two benchmarks.

\begin{figure}[h]
  \centering
\captionsetup{justification=raggedright,singlelinecheck=false}
    \includegraphics[width=0.49 \textwidth]{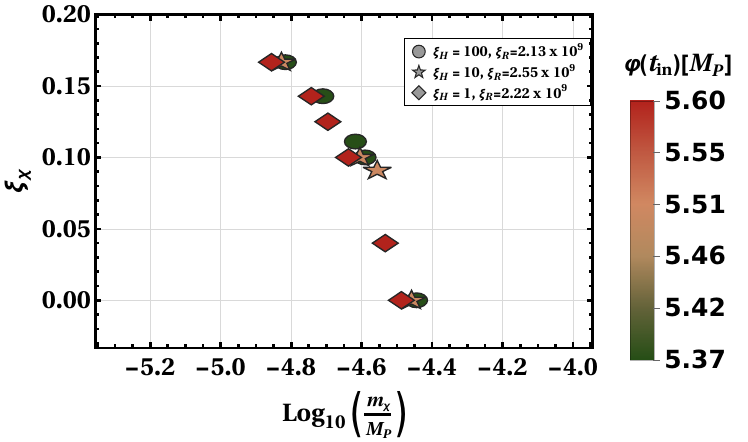}
    \includegraphics[width=0.49 \textwidth]{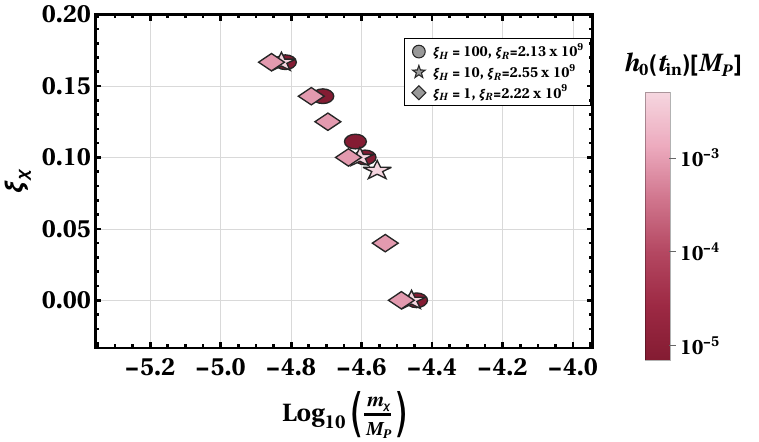}
\caption{DM parameter space ($\mx,\,\xi_\chi$) satisfying observables DM abundance $\omegax\simeq0.12$ for BPa($\circ$), BPb ($\star$), and BPc ($\diamond$). Left panel shows corresponding dependence on $\varphi(t_{\rm in})$ and right panel displays same scanned points with dependence on $h_0(t_{\rm in})$ represented by color-function bar.}\label{plot:dmparameter}
\end{figure}

Recall that in Fig.~\ref{relicplot}, we took $T_{\rm pre}\sim 5\times10^{14}$ GeV. We now find that the estimated $T_{\rm pre}$ for all three BPs lie within the same order of magnitude. Furthermore, we have found that preheating is completed within approximately 1.5--2 $e$-folds after the end of inflation, allowing $Y_\chi$ to reach its asymptotic plateau for $\mx\gtrsim\he$ before the linear theory breaks down. In Fig.~\ref{plot:dmparameter}, we show the scanned parameter space satisfying the observed DM relic abundance, $\omegax\simeq 0.12$, as a function of $\mx$ and $\xi_\chi$, along with its dependence on $\varphi(t_{\rm in})$ (left panel) and $h_0(t_{\rm in})$ (right panel), for all three BPs.
Here, we use the corresponding values of $T_{\rm pre}$ from Table~\ref{table:reheating}, providing a more complete version of Fig.~\ref{relicplot}.

Fig.~\ref{plot:dmparameter} illustrates, for the ballpark parameter space of Table~\ref{table:bp}, the $\mx\gtrsim 4\times 10^{-5} M_{\rm P}$ and $\xi_\chi < \frac{1}{6}$ to account for the observed relic density. The results also demonstrate that a finite parameter space exists where the observed DM relic abundance can be reproduced. More importantly, they show that preheating through the SM particle content alone can yield a $T_{\rm pre}$ compatible with the observed DM relic abundance, with the DM abundance reaching its asymptotic value before the completion of preheating, while the linear perturbative treatment remains valid throughout. For $\xi_H \lesssim 1$, preheating through the Goldstone bosons also becomes ineffective, and thermalization would instead proceed through perturbative reheating, which is beyond the scope of this paper.

\section{Summary and Outlook}
\label{sec:discu}
In this work, we have studied the cosmological gravitational production of a spectator scalar field $\chi$ as a DM candidate in a multifield $R^2$-Higgs inflationary framework.
After transforming from the Jordan to the Einstein frame, the spectator field acquires a time-dependent effective mass through its gravitational coupling to the background fields. We decomposed this effective mass into contributions arising from the effective potential, field-space curvature, the background-field evolution, and the Ricci-curvature contribution. This decomposition allows us to identify the physical origin of the particle-production mechanism. Focusing on $R^2$-like and mixed $R^2$-Higgs regime, we find that the behavior of the effective mass is strongly dependent on the non-minimal coupling $\xi_\chi$. For  $0 < \xi_\chi \lesssim \frac{1}{6}$, the effective mass squared becomes negative during inflation, causing the effective frequency squared to become negative for modes $\frac{k^2}{a^2} < |m_{\mathrm{eff},(h)}^2|$ with $m_{\mathrm{eff},(h)}^2<0$. The resulting tachyonic instability leads to exponential growth of the spectator-field mode functions these modes. After inflation, as long as $m_\chi \gtrsim H_e$,  the effective mass squared oscillates but remains positive, so the spectator field does not experience sustained parametric growth and the comoving number density $Y_\chi$ approaches an approximately constant value. For $m_\chi <H_e $, we find that the $Y_\chi$ keeps on increasing and does not freeze within a reasonable number of $e$-folding required for the validity of the linear approximation. We find also that, if $\xi_\chi>\frac{1}{6}$, no DM particle gets excited above the vacuum, which is true even for $m_\chi \gtrsim H_e$.

Using the mode functions obtained from the numerical evolution, we calculated the occupation number and the vacuum-subtracted comoving number density $Y_\chi$. For the sub-Hubble mass DM considered in this work, $Y_\chi$ reaches an asymptotic value shortly after the end of inflation and before the completion of preheating. Finally, we calculated the present-day relic abundance and studied its dependence on the spectator mass and the non-minimal coupling. We have seen that the relic abundance $\omegax$ decreases monotonically with increasing $\mx$ in the sub-Hubble mass regime, reflecting the exponential suppression of gravitational production for increasingly heavy spectator fields. Like for BP$b$, for the conformally coupled case $\xi_\chi=\frac{1}{6}$, the observed abundance $\omegax\simeq0.12$ can be reproduced up to $m_\chi\simeq5.0\,H_e \simeq3.5\times10^{13}\,\mathrm{GeV}$, while for $\xi_\chi=0$ the corresponding value is $m_\chi\simeq11.8 \,H_e\simeq8.4\times10^{13}\,\mathrm{GeV}$. Finally, scanning across all three benchmark points, we mapped out the two-dimensional parameter space $(\mx, \xi_\chi)$ consistent with the observed relic abundance (Fig.~\ref{plot:dmparameter}), finding a clear correlation between the allowed coupling $\xi_\chi$ and the initial field values $\varphi(t_{\rm in})$ and $h_0(t_{\rm in})$ that define each benchmark trajectory.

We then investigated the post-inflationary preheating dynamics for these benchmark points. The resulting preheating dynamics determines the energy density and temperature at the end of the preheating stage. In particular, we obtain $\tpre\simeq4.86\times10^{14}\,\mathrm{GeV},5.03\times10^{14}\,\mathrm{GeV},$ and $7.2\times10^{14}\,\mathrm{GeV}$ for BP$a$, BP$b$, and BP$c$, respectively. An important feature of our framework is that the preheating temperature is not treated as an independent free parameter; rather, it is determined by the underlying inflationary model parameters and the resulting background dynamics. This preheating temperature $T_{\rm pre}$, is used to fix the post-inflationary expansion history entering the abundance calculation. In addition, we also demonstrate that the $Y_\chi$ reaches its asymptotic flat region within $\mathcal{N} \sim 1.2$, before the end of preheating when the linear order theory breaks down.

Several directions remain open for future work. First, it would be interesting to consider more general spectator fields, including complex scalars, self-interacting fields, and higher-spin dark matter candidates like fermion \cite{Chung:2011ck,Ema:2019yrd}, vector \cite{Graham:2015rva}, within the same $R^2$-Higgs framework. Second, extending our linear analysis to include the backreaction of the produced $\chi$ quanta on the background and preheating dynamics would test the robustness of the asymptotic $Y_\chi$, particularly in the tachyonic regime $\xi_\chi\lesssim\frac{1}{6}$. Third, a systematic scan over the inflationary parameters $(\lambda,\xi_R,\xi_H)$, combined with current CMB constraints on $(n_s,r)$, could establish a direct connection between the inflationary parameter space and the gravitationally produced DM abundance.

\acknowledgments
AS thanks Bikash Kumar Acharya for the helpful discussion.

\bibliography{references.bib}
\end{document}